\documentclass[fleqn,usenatbib,useAMS]{mnras}
\usepackage{graphicx}	
\usepackage{amsmath}	
\usepackage{multicol}        
\usepackage{bm}		
\usepackage{pdflscape}	
\usepackage{latexsym}
\usepackage[authoryear]{natbib}
\usepackage{bm,nicefrac}
\usepackage{graphicx,mathtools,kantlipsum}
\usepackage{caption}
\usepackage[T1]{fontenc}
\usepackage{ae,aecompl}
\usepackage{newtxtext,newtxmath}
\def\kkps{K~km~s$^{-1}$}
\def\kps{km~s$^{-1}$}
\newcommand{\Msolar}{M$_{\odot}$}

\newcommand{\micro}{$\mu$m}
\newcommand{\simi}{$\sim$}
\newcommand{\as}{$''$}
\newcommand{\am}{$'$}
\newcommand{\htwo}{H\,{\sc ii}}

\newcommand{\cetno}{C$^{18}$O}
\newcommand{\tco}{$^{13}$CO}
\newcommand{\twco}{$^{12}$CO}
\newcommand{\Hii}{\textrm{H}\textsc{ ii}}
\title[CCC and HFSs in AFGL\,5180 and AFGL\,6366S]
{AFGL\,5180 and AFGL\,6366S: sites of hub-filament systems at the opposite edges of a filamentary cloud}  
\author[Maity et al.]
{A.~K. Maity$^{1,2}$\thanks{arupmaity@prl.res.in}, L.~K. Dewangan$^{1}$, N.~K. Bhadari$^{1,2}$, D.~K. Ojha$^{3}$, Z. Chen$^{4}$, \newauthor and Rakesh Pandey$^{1}$\\
$^{1}$Physical Research Laboratory, Navrangpura, Ahmedabad - 380 009, India.\\
$^{2}$Indian Institute of Technology Gandhinagar Palaj, Gandhinagar 382355, India. \\
$^{3}$Department of Astronomy and Astrophysics, Tata Institute of Fundamental Research, Homi Bhabha Road, Mumbai 400 005, India.\\
$^{4}$Purple Mountain Observatory, Chinese Academy of Sciences Nanjing 210033, PR China.}
\begin{document}
\date{ }
\pagerange{\pageref{firstpage}--\pageref{lastpage}} \pubyear{2023}
\maketitle
\label{firstpage}
\begin{abstract}
We present a multi-scale and multi-wavelength study to unveil massive star formation (MSF) processes around sites AFGL\,5180, and AFGL\,6366S, both hosting a Class\,{\sc II} 6.7\,GHz methanol maser emission. The radio continuum map at 8.46\,GHz reveals a small cluster of radio sources toward AFGL\,5180. Signatures of the early stages of MSF in our target sites are spatially seen at the opposite edges of a ﬁlamentary cloud (length {\simi}5\,pc), which is observed in the sub-millimeter dust continuum maps. Using the near-infrared photometric data, the spatial distribution of young stellar objects is found toward the entire ﬁlament, primarily clustered at its edges. The {\it getsf} utility on the {\it Herschel} far-infrared images reveals a hub-ﬁlament system (HFS) toward each target site. The analysis of the molecular line data, which benefits from large area coverage ({\simi}1{\rlap{$^\circ$}}\,\,\,$\times$ 1{\rlap{$^\circ$}}\,\,\,), detects two cloud components with a connection in both position and velocity space. 
This supports the scenario of a cloud-cloud collision (CCC) that occurred {\simi}1\,Myr ago.
The ﬁlamentary cloud, connecting AFGL\,5180 and AFGL\,6366S, seems spatially close to an {\htwo} region Sh2-247 excited by a massive O9.5 star. Based on the knowledge of various pressures exerted by the massive star on its surroundings, the impact of its energetic feedback on the ﬁlamentary cloud is found to be insigniﬁcant. Overall, our observational outcomes favor the possibility of the CCC scenario driving MSF and the formation of HFSs toward the target sites.
\end{abstract}
\section{Introduction}
\label{sec:intro}
Massive stars ($M$ $\gtrsim$ 8\,{\Msolar}) are known for their tremendous radiative and mechanical feedback, which allow them to play a vital role in the evolution of their host galaxies \citep[][and references therein]{Bressert_2012_new,tremblin_2014}. Despite their importance, the formation mechanism of massive stars is not yet fully understood \citep[e.g.,][]{zinnecker07,tan14,motte_2018}. In this relation, understanding the mass accumulation processes in massive star formation (MSF) is essential, which also includes the knowledge of the initial conditions of MSF. Hence, it requires extensive observational study of the surrounding environment involved in the early stages of MSF. The presence of massive stars is often evident through the detection of radio continuum emission from the H\,{\sc ii} regions \citep{Sharpless_1959} and/or the detection of Class\,II 6.7\,GHz methanol maser emission \citep[MME;][]{menten_1991,caswell_1995,walsh_1998}. 
Such proxies of MSF are commonly observed toward the compact hub or dense region \citep{schenider_2012,tige_2017,dewangan_hfs_2017,dewangan_hfs_2020}, which resides at the junction of several filaments \citep[i.e., hub-filament systems (HFSs);][]{myers_2009}. It is thought that  MSF starts inside hot molecular cores. As it evolves, it drives a hyper-compact (HC) H\,{\sc ii} region \citep[diameter ($d$) $\lesssim$ 0.05\,pc and electron density ($n_\mathrm{e}$) $\gtrsim$ 10$^5$\,cm$^{-3}$;][]{Kurtz_2005,Yang_2021} and then an ultra-compact (UC) H\,{\sc ii} region \citep[$d$ $\lesssim$ 0.1\,pc and $n_\mathrm{e}$ $\gtrsim$ 10$^4$\,cm$^{-3}$;][]{Churchwell_2002,Hoare_2007}. It is also believed that the 6.7\,GHz MMEs exclusively trace the early stages of MSF \citep{minier_2001,Breen_2013}. Such sites are very promising for probing the physical processes involved in the mass accumulation of MSF. In this context, the present paper deals with a target area hosting two potential massive star-forming regions, AFGL\,5180 and AFGL\,6366S, and each of them is associated with a 6.7\,GHz MME.

The site AFGL\,5180 is also known as G188.946$+$00.886 or IRAS 06058$+$2138  (hereafter, T1), while the region AFGL\,6366S is referred to as G189.030$+$0.784 or IRAS 06056$+$2131 (hereafter, T2). Various distances have been reported to these sites in the literature, which are 1.0–1.5\,kpc \citep{bik_2005}, 1.76\,kpc \citep{oh_2010,Mutie_2021}, 2.1\,kpc \citep{Reid_2009,Shimoikura_2013}, 2.2\,kpc \citep{koempe_1989}, and 3.5\,kpc \citep{moffat_1979,saito_2007}. In this work, we have adopted a distance of 1.5\,kpc, which is obtained from the examination of distances of Global Astrometric Interferometer for Astrophysics ({\it Gaia}) sources distributed toward the target sites (see Section~\ref{sec:phy_env} for more details). Previously, \citet{leistra_2006} also used the distance value of 1.5\,kpc as an average estimation between the proposed values of \citet{bik_2005} and \citet{koempe_1989}. The same distance value is also utilized in the other 
works \citep[e.g.,][]{carpenter_1993,devine_2008}.

Each of the target sites, T1 and T2 hosts one 6.7\,GHz MME with systemic velocities of 2.8 and 1.7\,{\kps}, respectively \citep{szymczak_2018}. 
Including the MMEs, several signposts of star formation (i.e., outflow activity, water maser, and young stellar objects (YSOs)) and dense clumps were observed toward both the sites \citep[e.g.,][]{koempe_1989,Davis_1998,ghosh_2000,klein_2005,wu_2010,Shimoikura_2013,Navarete_2015}. These sites are spatially found close to an {\htwo} region Sh2-247 (hereafter, S247) excited by a massive O9.5 star {\it LS V +21$^\circ$27} \citep{roman_2019}. The massive star is also known as ALS\,8736 or CGO\,115 (hereafter, cgo115).  

In the direction of the site T1, high-resolution near-infrared (NIR) images from the Integral Field Spectrograph SINFONI on UT4 (Yepun) of the VLT \citep[][]{Vasyunina_2010} and millimeter maps from the Atacama Large Millimeter/submillimeter Array (ALMA) have been examined \citep[e.g.,][]{Mutie_2021}. Ground-based NIR observations were also made toward T1 \citep{leistra_2006,devine_2008}, and the age of the NIR cluster toward T1 was reported to be $\sim$2.5\,Myr \citep{devine_2008}. Using the NIR spectroscopic study, \citet{Vasyunina_2010} obtained a similar age of 1-3\,Myr for T1. Previous results also suggested that T1 is younger than T2 \citep{kurtz_1994,ghosh_2000}.  
Using the James Clerk Maxwell Telescope (JCMT) 850\,{\micro} dust continuum map, both the sites T1 and T2 were found to be connected by a filamentary feature \citep[see Figure~1 around IRAS~06056+2131 and IRAS~06058+2138 in][]{klein_2005}. Molecular emission were also detected toward this filamentary feature \citep[see Figures~1b and 1c in][]{Shimoikura_2013}. However, no study is yet carried out to explore the role of this filamentary feature in star-forming activity in both sites. 
\citet{carpenter_1995a,carpenter_1995b} proposed a triggered star formation scenario for the Gemini OB1 molecular cloud complex, including our selected target sites. Based on the molecular line data, \citet{Shimoikura_2013} reported two velocity components around [$-$3, 5] and [5, 13]\,{\kps} (see Figure 3b in their paper) and suggested the applicability of cloud-cloud collision (CCC) process in both the sites for explaining the observed star formation activities. 

Our selected targets host early phases of MSF and are connected with a filamentary feature as traced in previous studies of dust continuum and molecular line data. However, despite numerous studies on the target sites, no attempt has been made to study embedded filaments and their role in the mass accumulation processes in MSF. In this relation, this paper examines the applicability of the previously proposed CCC scenario and the role of filament in star formation processes. It also attempts to understand the impact of the O-type star cgo115 on the sites T1 and T2. An extensive analysis of the multi-scale and multi-wavelength data sets (see Section~\ref{sec:obser} for details) has been performed in this relation. Embedded features are studied at small-scales ({\simi}0.01\,pc) to large-scales ({\simi}1\,pc) using several continuum images covering NIR to radio wavelengths. Additionally, molecular line data were employed to examine the kinematics of molecular gas and their spatial distribution.

The paper is arranged as follows. Section~\ref{sec:obser} describes the details of the data utilized in this work. The observational results are presented in Section~\ref{sec:results} and discussed in Section~\ref{sec:dis}. Section~\ref{sec:conc} summarizes our main findings.
\section{Data sets}
\label{sec:obser}
Multi-scale and multi-wavelength data sets were collected from various surveys for our selected target area of 19{\rlap{\am}.58} $\times$ 18{\rlap{\am}.33} (centred at {{\it l}} = 188{\rlap{$^\circ$}.~98}, {\it b} = 0\rlap{$^\circ$}.~83).  
This work utilized the {\it Gaia} Early Data Release 3 (EDR3) to estimate the distance of our target source \citep{Gaia_Collaboration_2021,Bailer_Jones_2021}. A brief description of all other data sets analyzed in this work is presented below.
\subsection{NIR data}
Deep NIR photometric H and K band magnitudes of point-like sources were obtained from the UKIDSS-Galactic Plane Survey \citep[GPS;][]{Lawrence_2007} and the Two Micron All Sky Survey \citep[2MASS;][]{Skrutskie_2006}. We also collected the UKIDSS-GPS K band image for our target area. The UKIDSS-GPS observations (resolution $\sim$0\rlap{\as}.~8) were made using Wide Field Camera \citep[WFCAM;][]{Casali_2007} mounted on the 3.8\,m UK Infra-Red Telescope. We selected only those sources which have photometric errors of 0.1\,mag, or less in both the bands. The {\it Spitzer} IRAC 3.6 and 4.5\,{\micro} images (resolution of {\simi}2{\as}) and photometric magnitudes of point-like sources were collected from the Warm-{\it Spitzer} Glimpse~360 \citep{Whitney_2011} survey. The point-like sources with a photometric magnitude error of 0.2\,mag, or less in the IRAC 3.6 and 4.5\,{\micro} bands were considered for further analysis. Additionally, we obtained the Hubble Space Telescope (HST) pipeline-calibrated NIR imaging data (resolution $\sim$0\rlap{\as}. 1 --0\rlap{\as}. 2) in the J (WFC3/F110W), H (WFC3/F160W) wide band filters, and in the [Fe\,{\sc ii}] (WFC3/F164N) narrow band filter. These images are only available for site T1.
\subsection{H$_{\mathrm{2}}$ narrow band image}
Our entire selected target area was covered in the survey of extended H$_{\mathrm{2}}$ emission \citep{Navarete_2015}, which was carried out (with an average seeing of $\sim$0\rlap{\as}. 7) using the Wide-field InfraRed Camera (WIRCam) mounted on the Canada-France-Hawaii Telescope (CFHT). We used the continuum-subtracted H$_{\mathrm{2}}$ ($\lambda$ = 2.122\,{\micro}, $\Delta\lambda$ = 0.032\,{\micro}) image from this survey \citep[for more details, see][]{Navarete_2015}.
\subsection{Dust continuum data}
We obtained far-infrared (FIR) and sub-millimeter dust continuum images (70--500\,{\micro}) from the {\it Herschel{\footnote[1]{{\it Herschel} is an ESA space observatory with science instruments provided by European-led Principal Investigator consortia and with important participation from NASA.}}} Space Observatory data archives, which were collected as a part of {\it Herschel} Infrared Galactic Plane Survey \citep[Hi-GAL;][]{Molinari10a}. The resolution of the {\it Herschel} images at 70, 160, 250, 350, and 500\,{\micro} are about 6{\as}, 12{\as}, 18{\as}, 25{\as}, and 37{\as}, respectively. 

We obtained the {\it Planck} 353\,GHz or 850\,{\micro} dust continuum polarization data, comprising of Stokes I, Q, and U maps (resolution {\simi}5{\am}) for our entire target area \citep{Planck2014}. These data were extracted from the {\it Planck} Public Data Release 3 of Multiple Frequency Cutout Visualization (PR3 Full Mission Map with PCCS2 Catalog). Results derived using the {\it Planck} polarization data are presented in APPENDIX~\ref{sec:posbfield}.

The high-resolution ALMA 1.3\,mm dust continuum images (project id: 2015.1.01454.S, PI: Zhang, Yichen) at two different resolutions (i.e., 0\rlap{\as}. 18 $\times$ 0\rlap{\as}. 28 and 0\rlap{\as}. 63 $\times$ 1\rlap{\as}. 23) were collected toward the site T1 from the ALMA FITS Archive. Note that no ALMA observations were made toward site T2. An analysis of the high-resolution NIR images (from HST and UKIDSS) and the ALMA 1.3\,mm dust continuum map can be found in APPENDIX~\ref{sec:ALMA_res}.
%
\subsection{Molecular CO line data}
To explore the gas kinematics, we examined the $^{12}$CO(J = 1--0; rest frequency ($\nu_{0}$) $\sim$115.27\,GHz), $^{13}$CO(J = 1--0, $\nu_{0}$ $\sim$110.20\,GHz) and C$^{18}$O(J = 1--0, $\nu_{0}$ $\sim$109.78\,GHz) line data toward our selected target area. These molecular line data (beam size $\sim$55$''$) were taken as a part of the Milky Way Imaging Scroll Painting (MWISP\footnote[1]{http://english.dlh.pmo.cas.cn/ic/in/}) project. This project is an ongoing northern Galactic plane $^{12}$CO(J = 1--0)/$^{13}$CO(J = 1--0)/C$^{18}$O(J = 1--0) survey led by the Purple Mountain Observatory (PMO), using the 13.7\,m millimeter-wavelength telescope\footnote[2]{http://www.radioast.nsdc.cn/mwisp.php} at Delingha, China. The telescope mapped an area of 0{\rlap{$^\circ$}.~5} $\times$ 0{\rlap{$^\circ$}.~5} using position-switch On-The-Fly \citep[OTF; see][]{Sun_2018} mode with a pointing accuracy better than 5{\as}. The total bandwidth of 1\,GHz, with a frequency interval of 61\,kHz, provides 16,384 channels which result in a velocity separation of {\simi}0.16\,{\kps} for {\twco} and {\simi}0.17\,{\kps} for {\tco} and {\cetno}. The achieved rms noise level for {\twco} is {\simi}0.5\,K at the channel width of 0.16\,{\kps} and for {\tco} and {\cetno} is {\simi}0.3\,K at the channel width of 0.17\,{\kps}. These molecular data were reduced using the GILDAS software\footnote[3]{http://ascl.net/1305.010 or http://www.iram.fr/IRAMFR/GILDAS} \citep{Pety_2005}. More about the telescope and receiver employed in the MWISP project can be found in \citet{Su_2019} and references therein.
\subsection{Radio continuum data}
The 1.4\,GHz radio continuum map for our entire target area was collected from the NRAO VLA Sky Survey \citep[NVSS;][]{NVSS}, and the radio map has the resolution and sensitivity of {\simi}45{\as} and {\simi}0.45\,mJy\,beam$^{-1}$, respectively. We also obtained the processed radio continuum map at 8.46\,GHz (beam size $\sim$3\rlap{\as}. 1 $\times$ 2\rlap{\as}. 3; rms $\sim$36.8\,$\mu$Jy\,beam$^{-1}$) from the NRAO Very Large Array Archive Survey (NVAS) toward T1 only. 
\section{Results}
\label{sec:results}
\subsection{Physical environment of the target sites T1 and T2}
\label{sec:phy_env}
\subsubsection{Elongated filament}
\label{sec:phy_env1}
\begin{figure*}
\includegraphics[width= \textwidth]{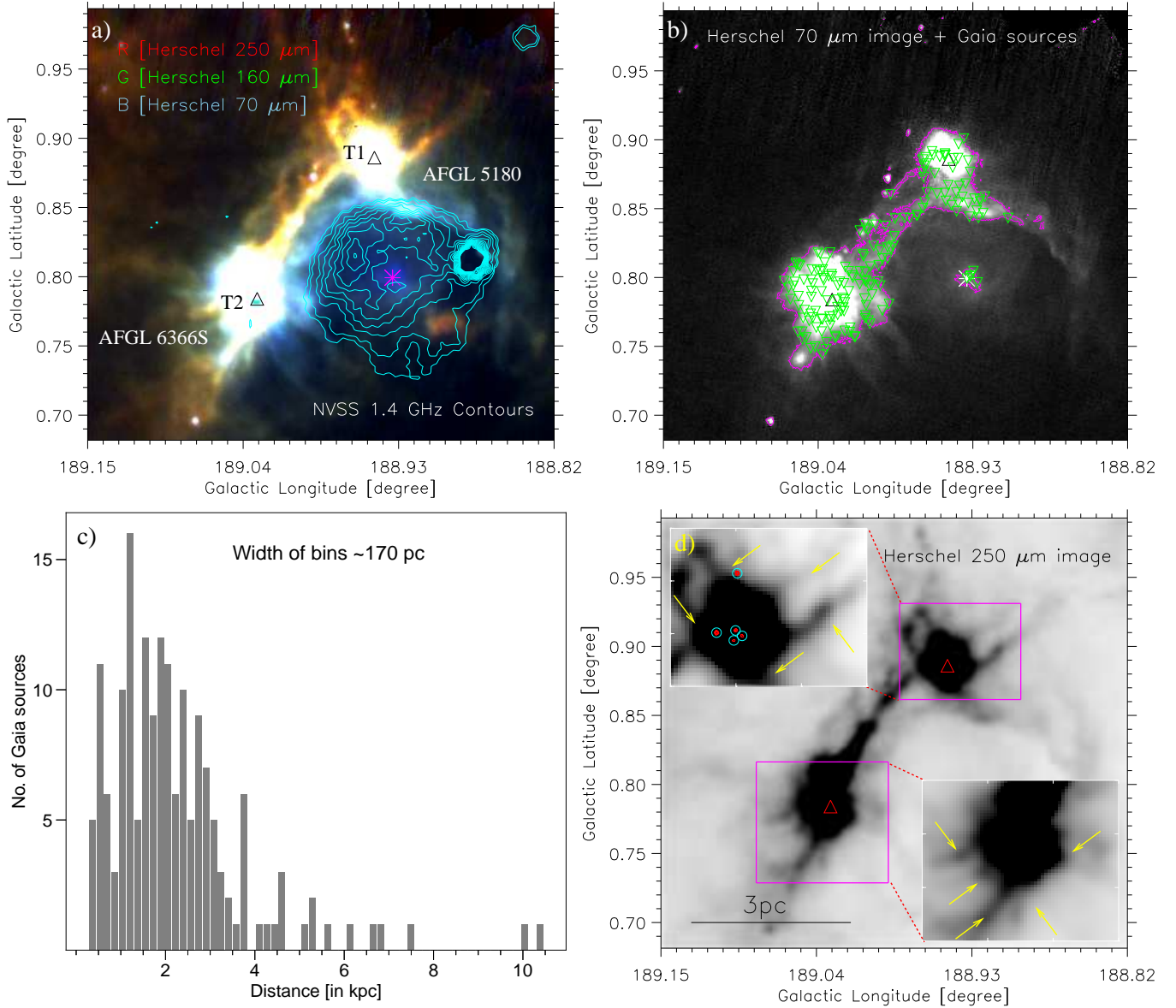}
\caption{a) Overlay of the NVSS 1.4\,GHz radio continuum emission contours on a three-color composite map of an area hosting AFGL\,5180 and AFGL\,6366S. The color composite map consists of the {\it Herschel} 250\,$\mu$m image (in red), 160\,$\mu$m image (in green), and 70\,$\mu$m image (in blue). The NVSS radio continuum contours are plotted with levels of 3$\sigma$ and 5$\sigma$ to 80$\sigma$ with an interval of 5$\sigma$, where 1$\sigma$ $\sim$0.45\,mJy\,beam$^{-1}$. b) The panel shows the positions of the {\it Gaia} sources overlaid on the {\it Herschel} image at 70\,$\mu$m (see upside down green triangles). The {\it Gaia} sources are mainly selected toward an elongated feature, which is indicated by a contour (in magenta) with a level of 0.06\,Jy\,pixel$^{-1}$. c) The distance distribution of the selected {\it Gaia} sources. d) The panel shows the inverted grayscale {\it Herschel} image at 250\,{\micro}. The insets on the top-left and bottom-right present zoomed-in views of the sites AFGL\,5180 (or T1) and AFGL\,6366S (or T2), respectively, using the inverted grayscale {\it Herschel} image at 250\,{\micro}. In the direction of T1, the NVAS 8.46\,GHz radio continuum contours at 120, 240, and 360\,$\mu$Jy\,beam$^{-1}$ (where, 1$\sigma$ {\simi}36.8\,$\mu$Jy~beam$^{-1}$) are shown in red. The peak positions of these radio continuum emissions are indicated by cyan circles (see the inset on the top-left). Arrows highlight sub-filaments toward both target sites. A scale bar of 3\,pc is shown for a distance of 1.5\,kpc. In panels ``a'', ``b'' and ``d'' triangles indicate the positions of the 6.7\,GHz MMEs. In ``a'' and ``b'', the asterisk indicates the position of an O-type star, which is referred to as cgo115.} 
\label{fig1}
\end{figure*}

Figure~\ref{fig1}a displays a {\it Herschel} three-color composite image (red: 250\,$\mu$m, green: 160\,$\mu$m, and blue: 70\,$\mu$m), which is overlaid with the NVSS\,1.4 GHz radio continuum emission contours and the position of the previously known massive star, cgo115. The positions of two 6.7\,GHz MMEs are also indicated in Figure~\ref{fig1}a (see triangles). An elongated filamentary structure is the most prominent feature in the color composite map, which has a dumbbell-like appearance. Each end of the filamentary structure is associated with at least one earlier reported star-forming site, where one 6.7\,GHz MME is also traced. The ionized emission outlined through the NVSS 1.4\,GHz radio continuum contours shows the spatial extension of the H\,{\sc ii} region, S247. The massive star, cgo115, seems to be located at the center of the radio continuum emission (see an asterisk in Figure~\ref{fig1}a). The NVSS 1.4\,GHz radio continuum emission (1\,$\sigma$ $\sim$0.45\,mJy\,beam$^{-1}$) is not detected toward the filamentary structure, including both the star-forming sites. Figure~\ref{fig1}b shows the {\it Herschel} 70\,$\mu$m image overlaid with the positions of the {\it Gaia} EDR3 sources \citep{Gaia_Collaboration_2021}. 170 {\it Gaia} sources are selected toward areas enclosed by the 70\,$\mu$m emission contour (see Figure~\ref{fig1}b). The photogeometric distances (``rpgeo'') specified in \citet{Bailer_Jones_2021} are utilized for the selected {\it Gaia} EDR3 sources. Figure~\ref{fig1}c presents the distance distribution of these {\it Gaia} sources. Distribution of these {\it Gaia} sources peaks around 1.5\,kpc, which is consistent with the earlier adopted distance for the site T1 \citep{leistra_2006,devine_2008}. The angular separations of the 6.7\,GHz MMEs in T1 and T2 with the position of the massive star, cgo115, are determined to be about 5\rlap{\am}.3 (2.3\,pc) and 5\rlap{\am}.9 (2.6\,pc), respectively. The angular separation between these two MMEs is about 7\rlap{\am}.9 (3.4\,pc). We present the inverted grayscale {\it Herschel} 250\,$\mu$m image in Figure~\ref{fig1}d. It reveals an elongated filamentary structure (of length {\simi}4\,pc), hosting T1 and T2 at the opposite edges of the filament.
\subsubsection{Hub-filament systems}
\label{fil}
Apart from the elongated filamentary structure, we notice a HFS toward both the sites AFGL\,5180 and AFGL\,6366S, where several small scale filaments (i.e., sub-filaments; length {\simi}1\,pc) appear to direct toward the central regions. To highlight HFSs, zoomed-in views of T1 and T2 are presented using the inverted grayscale FIR map at {\it Herschel} 250\,$\mu$m (see insets in Figure~\ref{fig1}d). The NVAS 8.46\,GHz radio continuum map is available toward site T1. However, site T2 is not covered by the NVAS survey. One can notice that the NVAS survey has better sensitivity than the NVSS survey. In Figure~\ref{fig1}d, the position of the radio continuum sources detected in the NVAS 8.46\,GHz radio continuum map are presented in the inset on the top-left, indicating the presence of a small cluster of radio sources (or massive stars) toward AFGL\,5180.

In order to further examine the embedded structures, we utilized a newly developed algorithm, {\it getsf} \citep{getsf_2022} on the {\it Herschel} 160\,$\mu$m image (resolution {\simi}12{\as}). Use of {\it getsf} disintegrates the astronomical image into its structural components (i.e., sources and filaments) and splits these components from each other and their backgrounds. One can find more details about this algorithm in \citet{getsf_2022}. The maximum size of the structural elements (i.e., filaments and sources), the angular resolution of the image, and the source distance are the necessary inputs for the algorithm to function. Here, the filamentary skeletons in the {\it Herschel} 160\,{\micro} image are extracted with the maximum source and filament sizes of 20{\as} and 105{\as}, respectively, and a distance of 1.5\,kpc. The outcome of the {\it getsf} utility at the scale of 12{\as} is presented in Figure~\ref{fig2}a (see filamentary skeletons in red). Several 
sub-filaments are identified in the direction of both the sites T1 and T2, which are highlighted on the {\it Herschel} 160\,{\micro} image in Figure~\ref{fig2}a. 
The identified sub-filaments make a junction toward the central part of T1 and T2, which are associated with the 6.7\,GHz MMEs (see green triangles in Figure~\ref{fig2}a). Therefore, MSF activities are exclusively found at the junctions of several sub-filaments. We have highlighted two distinct structures (i.e., a curved feature F1 and an elongated filament F2)
in Figure~\ref{fig2}a. The curved feature F1 is observed at the boundary of the {\htwo} region traced in the NVSS 1.4\,GHz radio continuum emission map. The sites T1 and T2 are seen at opposite edges of filament F2. Hence, the presence of HFSs at opposite edges of an elongated filament is a new picture concerning the target sites.
\subsubsection{Feedback of the O-type star} 
\label{sec:pre}
\begin{figure}
\includegraphics[width= 7 cm]{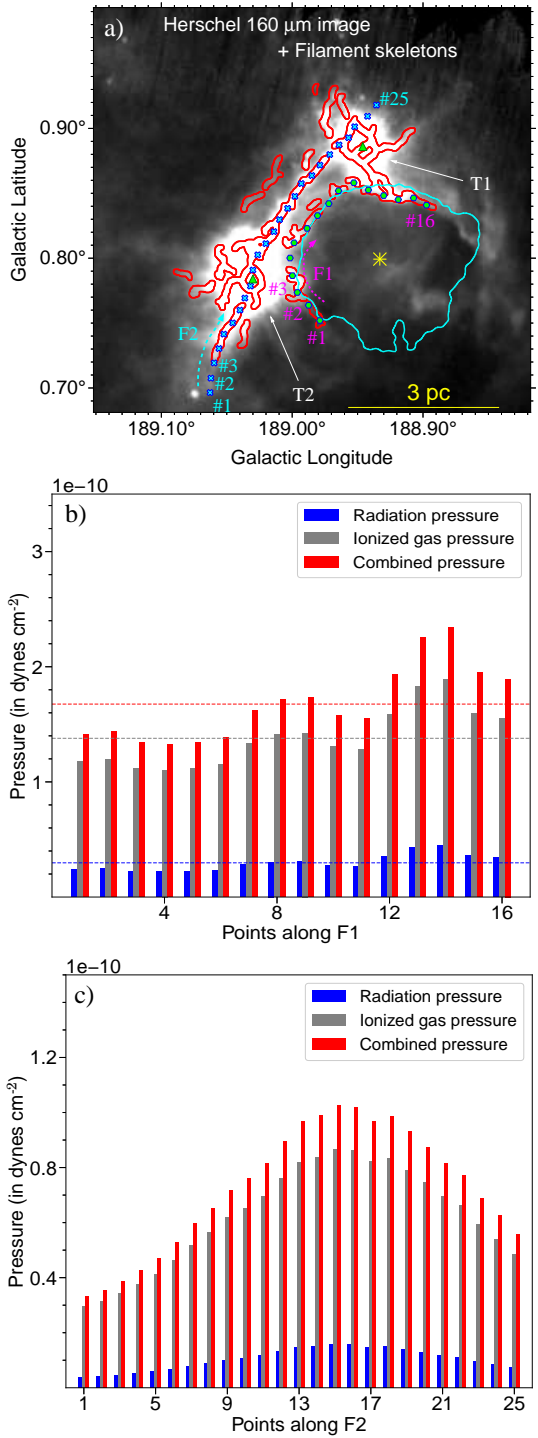}
\caption{ a) The panel shows the {\it Herschel} 160\,{\micro} continuum image overlaid with the NVSS 1.4\,GHz radio continuum emission contour (in cyan) at 1.35\,mJy\,beam$^{-1}$ and the filament skeletons (in red) identified using the {\it getsf} utility. The panel indicates two distinct structures: a curved feature F1 and an elongated filament F2. The filled hexagons and crosses mark several positions along F1 and F2 (see Section~\ref{sec:pre} for more details), respectively. These positions are also labeled in the panel. The asterisk and the triangles are the same as presented in Figure~\ref{fig1}a. A scale bar representing 3\,pc is also shown. b) The variation of radiation pressures (P$_{\mathrm{rad}}$; in blue), ionized gas pressures (P$_{\Hii}$; in gray), and combined pressures (i.e., P$_{\Hii}$ + P$_{\mathrm{rad}}$ + P$_{\mathrm{wind}}$; in red) computed at selected positions along F1, exerted by cgo115. The horizontal dashed lines indicate the average values of the pressure distributions. c) Same as Figure~\ref{fig2}b, but it is shown for the positions along F2.}
\label{fig2}
\end{figure}

To assess the influence of cgo115 (see a yellow asterisk in Figure~\ref{fig2}a) on the curved feature F1 and the filament F2, we have calculated the strength of different pressure components driven by the massive O-type star on these structures. The equations leading to thermal pressure of the ionized gas (P$_{\Hii}$), radiation pressure (P$_{\mathrm{rad}}$), and stellar wind ram pressure (P$_{\mathrm{wind}}$) are obtained from {\citet{{Bressert_2012_new}}} and given below,
\begin{equation}
P_{\Hii} = \mu m_{\mathrm{H}} c_{\mathrm{s}}^2\, \left(\sqrt{3N_{\mathrm{uv}}\over 4\pi\,\alpha_{\mathrm{B}}\, D_{\mathrm{s}}^3}\right);
\end{equation}
\vspace{-0.2 cm}
\begin{equation}
P_{\mathrm{rad}} = L_{\mathrm{bol}}/ 4\pi c D_{\mathrm{s}}^2; 
\end{equation}
\vspace{-0.5 cm}
\begin{equation}
P_{\mathrm{wind}} = \dot{M}_{\mathrm{w}} V_{\mathrm{w}} / 4 \pi D_{\mathrm{s}}^2. 
\end{equation}
Here, $N_\mathrm{uv}$ is the number of Lyman continuum photons emitted per second by the ionizing source. The sound speed in the ionized region is $c_{\mathrm{s}}$ = 11\,{\kps} \citep{Bisbas_2009} and the radiative recombination coefficient is $\alpha_{\mathrm{B}}$ = 2.6 $\times$ 10$^{-13}$\,cm$^{3}$\,s$^{-1}$ \citep{Kwan_1997}, assuming the temperature to be around 10$^4$\,K. The mean molecular weight in the ionized gas is $\mu$ = 0.678 \citep{Bisbas_2009}, and $m_{\mathrm{H}}$ indicates the mass of the hydrogen atom. $\dot{M}_{\mathrm{w}}$,  $V_{\mathrm{w}}$ and $L_{\mathrm{bol}}$ are respectively the mass-loss rate through stellar wind, the wind velocity, and the bolometric luminosity of the ionizing source, cgo115. The speed of light ($c$) is adopted to be 2.9979 $\times$ 10$^{8}$\,m\,s$^{-1}$. $D_{\mathrm{s}}$ is the projected distance between the point of our interest and the massive star. For a O9.5V star, we can adopt $L_{\mathrm{bol}}$ $\approx$ 66070\,L$_{\odot}$ \citep{Panagia_1973}, $\dot{M}_{\mathrm{w}}$ $\approx$ 1.58 $\times$ 10$^{-9}$\,M$_{\odot}$\,yr$^{-1}$ \citep{Marcolino_2009}, $V_{\mathrm{w}}$ $\approx$ 1500\,{\kps} \citep{Martins_2017} and $N_{\mathrm{uv}}$ $\approx$ 1.2 $\times$ 10$^{48}$\,s$^{-1}$ \citep{Panagia_1973}. 
\par
To calculate the pressure components mentioned above, 16 positions are selected on the structure F1 (see filled hexagons in Figure~\ref{fig2}a). The first three positions (i.e., \#1, \#2, and \#3) and the last position (i.e., \#16) are mentioned to follow the order of numbering of the positions. For the same purpose, 25 positions are chosen along the filament F2 (see filled crosses in Figure~\ref{fig2}a). The pressure components P$_{\mathrm{rad}}$, P$_{\Hii}$, and the combined pressure (i.e., P$_{\mathrm{wind}}$ $+$ P$_{\mathrm{rad}}$ $+$ P$_{\Hii}$) are displayed in Figures~\ref{fig2}b and~\ref{fig2}c for positions along F1 and F2, respectively. Hence, we find that along F1, all pressure components fluctuate around their mean values (see horizontal lines in Figure~\ref{fig2}b). For F2, the closest position has the highest pressure in all forms, and all pressure components fall on either side as their distances from the massive star increase. Overall, ionized gas pressure dominates over the other pressure components for our region of interest. The wind pressures present a negligible contribution of about 10$^{-16}$\,dynes\,cm$^{-2}$ in the total pressure values for both F1 and F2. From Figure~\ref{fig2}c, we can infer that the combined pressure values excluding P$_{\Hii}$ (i.e., the total of P$_{\mathrm{wind}}$ and P$_{\mathrm{rad}}$) along F2 are about 10$^{-11}$\,dynes\,cm$^{-2}$. It is important to note that the filaments and the massive star are assumed to lie on the same plane as of sky. Thus, our estimated distances represent the lower limit that leads to the upper limit in the estimated pressure values if the projection effect is considered.
\subsection{Study of embedded protostars}
\label{sec:yso}
\begin{figure}
\includegraphics[width= 7 cm]{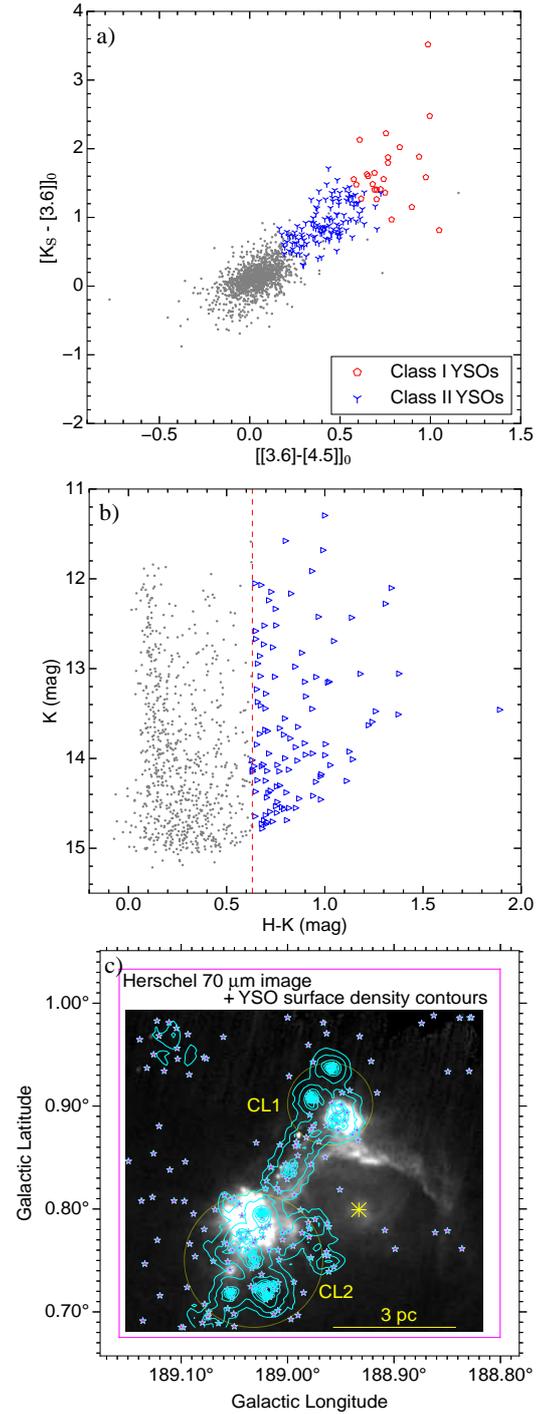}
\caption{a) Dereddened color-color ([[3.6]$-$[4.5]]$_0$ Vs. [K$-$[3.6]]$_0$ ) diagram of point-like sources distributed toward our selected target area. Class\,{\sc I} and Class\,{\sc II} YSOs are indicated by pentagons (in red) and tri-down symbols (in blue), respectively. b) The panel shows the NIR color-magnitude (H$-$K Vs. K) diagram of point-like sources (see Section~\ref{sec:yso} for details). Color-excess sources are highlighted by triangle-right symbols (in blue). 
c) Overlay of the positions of the selected YSO candidates (see magenta stars) on the {\it Herschel} 70\,{\micro} image. The surface density contours of YSOs are shown by cyan contours, and their levels are at 15, 25, 50, 75, 100, 125, 150, 175, 200, 225, 250, 275, and 300\,YSOs\,pc$^{-2}$. The circles (in yellow) highlight the clusters of YSOs, which are found toward the edges of the filamentary structure.
The scale bar and the asterisk are the same as Figure~\ref{fig2}a. The magenta rectangle shows the area of the molecular data displayed in Figure~\ref{fig5}.}
\label{fig3}
\end{figure}

The detection of YSOs has been adopted as a reliable tool to trace ongoing star formation activity in a given star-forming site. 
Therefore, the procedure of \citet{Gutermuth_2009} is followed to identify the embedded YSOs using the photometric data of point-like objects in the area as shown in Figure~\ref {fig1}a. A dereddened color-color diagram (i.e., [[3.6]$-$[4.5]]$_{0}$ Vs. [K$-$[3.6]]$_{0}$) is presented in Figure~\ref{fig3}a, revealing 25 Class\,I YSOs and 128 Class\,II YSOs. Additional YSOs are also selected using the NIR color-magnitude diagram (i.e., H$-$K Vs. K) as displayed in Figure~\ref{fig3}b. Following the SQL conditions presented in \citet{Lucas_2008}, the reliable photometric data in the H and K bands were obtained from the UKIDSS-GPS survey. In general, bright sources are saturated in the UKIDSS-GPS survey. Hence, we also utilized the 2MASS photometric data in the H and K bands. The color-magnitude analysis of a nearby control ﬁeld provides a color condition of H $-$ K $>$ 0.63\,mag, allowing us to find 109 more YSO candidates. Altogether, 208 YSO candidates are selected after considering the common sources extracted from these two schemes. In Figure~\ref{fig3}c, the positions of these selected YSOs are overlaid on the {\it Herschel} 70\,{\micro} image.

The nearest neighbor (NN) surface density analysis is a useful tool to assess groups/clusters of the YSOs in star-forming regions \citep[see, e.g., ][]{Casertano_1985,Bressert_2010,Dewangan_2017S237,Bhadari_2020}. We have also adopted this technique to compute the surface density map of YSOs in our target area. The target area, as shown in Figure~\ref{fig1}a is divided into 100 $\times$ 100 grid lines, corresponding to the separations of 11\rlap.$''$7 and 11\rlap.$''$1 along the Galactic longitude and latitude, respectively. Considering d = 1.5\,kpc and NN = 5, the surface density map of YSOs is produced. In Figure~\ref{fig3}c, {\it Herschel} 160\,{\micro} image is overplotted with YSOs surface density contours at the levels of 15 -- 300\,YSOs\,pc$^{-2}$. Clusters of YSOs CL1 and CL2 are seen toward the sites T1 and T2, respectively (see yellow dotted circles in Figure~\ref{fig3}c). Noticeable YSOs are also traced toward the central part of the elongated filament. Hence, our analysis reveals the ongoing star formation activities toward the entire filament F2, and the clusters of YSOs are mainly found toward the edges of the filament. 
\subsection{{\it Spitzer} ratio map and continuum-subtracted H$_\mathrm 2$ emission}
\label{ratio_map}
The {\it Spitzer} ratio map (i.e., 4.5\,{\micro}/3.6\,{\micro}) and the narrow band H$_2$ ($\nu=1-0$ S(1), 2.12\,{\micro}) emission are very useful to trace the energetic feedback of massive stars and/or star formation activities in a given molecular cloud \cite[see][]{povich_2007,Dewangan_2017S237,pandey_2020}. A detailed description of the H$_2$ emission in massive star-forming sites M~17 UC1 – IRS5 can be found in \citet{Chen_2015}. In this context, Figure~\ref{fig4}a presents the {\it Spitzer} ratio map for our target area. Bright regions in the ratio map present an excess of 4.5\,{\micro} emission over 3.6\,{\micro} emission, while the exact inverse condition is true for the dark regions. The continuum-subtracted H$_2$ ($\nu=1-0$ S(1)) map is produced using a similar approach discussed in \citet{Navarete_2015}. Figure~\ref{fig4}b presents the continuum-subtracted H$_2$ ($\nu=1-0$ S(1)) map for a region highlighted with a magenta rectangle in Figure~\ref{fig4}a.

The {\it Spitzer} 4.5\,{\micro} band includes contributions from H$_2$ ($\nu=0-0$ S(9), 4.693\,{\micro}) and H\,{\sc i} Br-$\alpha$ recombination line emission (4.05\,{\micro}). On the other side, {\it Spitzer} 3.6\,{\micro} band covers far-ultraviolet (FUV) heated polycyclic aromatic hydrocarbon (PAH) 3.3\,{\micro} emission from C$-$H vibrational stretching mode \citep{Allamandola_1989,drain_2003,Tielens_2008} and H$_2$ ($\nu=1-0$ O(5), 3.234\,{\micro}). Due to an excess of 4.5\,{\micro} over 3.6\,{\micro} emission in the ratio map, bright regions are found toward the entire elongated filament feature (including the sites T1 and T2; see Figure~\ref{fig4}a), where the extended radio continuum emission is absent. In Figure~\ref{fig4}b, the H$_2$ ($\nu=1-0$ S(1)) emission is also seen toward the filament. Hence, Figures~\ref{fig4}a and~\ref{fig4}b support the presence of outflow activities toward the entire filament (including T1 and T2), indicating the ongoing star formation activities (see also clusters of YSOs in Section~\ref{sec:yso}). Dark features in the ratio map are traced at the periphery of the H\,{\sc ii} region, S247 (see also the curved feature F1), showing an excess of PAH-dominated 3.3\,{\micro} emission. A similar characteristic is also traced around the Galactic {\htwo} region S305 using the {\it Spitzer} ratio map \citep[e.g.,][]{dewangan2020_S305,Bhadari_2021}. Therefore, the dark regions in the ratio map may display the photo-dissociation regions (PDRs), suggesting the influence of the massive star on its surrounding. 
\subsection{{\it Herschel} column density and dust temperature maps}
\label{mapping}
The multi-wavelength {\it Herschel} images allow the computation of the column density ($N(\mathrm{H_2})$) and the dust temperature ($T_\mathrm{d}$) maps for a region indicated by a green rectangle in Figure~\ref{fig4}a. A pixel-by-pixel spectral energy distribution (SED) fitting with a modified blackbody spectra provides the desired physical parameters (i.e., $N(\mathrm{H_2})$ and $T_\mathrm{d}$) at each pixel. Despite the availability of the {\it Herschel} 70\,{\micro} image, this has been excluded from the SED fitting because it generally contains contamination from ultraviolet (UV) heated dust of much higher temperatures. We utilized the tool {\it hires} \citep{getsf_2022} to produce the high-resolution column density and dust temperature maps. The basic concept behind these high-resolution maps is the addition of higher-resolution information to the lower-resolution images as differential terms \citep[see Equation A.1 in][]{Palmeirim_2013}. More details can be found in \citet{getsf_2022}. The use of the {\it Herschel} images (i.e., 70--500\,{\micro}) in {\it hires} provides up to the best resolution available among the input images (i.e., {\simi}6{\as} for {\it Herschel} 70\,{\micro} image). However, the resulting column density and dust temperature maps of 6{\as} resolution are noisy, which may be due to insigniﬁcant features present in the {\it Herschel} 70\,{\micro} image. Hence, we have utilized the column density and dust temperature maps of resolution {\simi}12{\as} in this work, which are presented in Figures~\ref{fig4}c and~\ref{fig4}d, respectively. In the direction of the HFSs (see arrows toward T1 and T2 in Figure~\ref{fig4}c), the central hubs or junctions of the sub-filaments hosting the 6.7 GHz MMEs are found with higher column densities ($N(\mathrm{H_2})$ {\simi}10$^{23}$\,cm$^{-2}$). The central hubs are traced with higher dust temperatures of 22-26\,K (see Figure~\ref{fig4}d). Figure~\ref{fig4}e displays a two-color composite image using the column density map (in red) and the dust temperature map (in cyan). In Figure~\ref{fig4}e, the filament F2 and the curved feature F1 are clearly distinguished in the {\it Herschel} temperature map (see also Figure~\ref{fig4}d). The curved feature F1 shows a relatively higher dust temperature than the filament F2 except for its edges. In other words, the elongated filament F2 has a higher column density and lower dust temperature, but the curved structure F1 has a higher dust temperature and lower column density. 
\begin{table*}
\centering
\captionsetup{justification=centering,margin=5cm}
\caption{ List of the physical parameters of the detected {\it Herschel} clumps. $R_\mathrm{clump}$ is the effective radius and $[N(\mathrm{H_2})]_{\mathrm{peak}}$ presents the peak column density. $[T_\mathrm{d}]_\mathrm{peak}$ and $[T_\mathrm{d}]_\mathrm{mean}$ stand for the peak and mean dust temperature, respectively.}
\label{tab1}
\begin{tabular}{cccccccc}
\hline 
  ID  &  {\it l}         &  {\it b}       & $R_\mathrm{clump}$        &  $M_\mathrm{clump}$	& $[N(\mathrm{H_2})]_{\mathrm{peak}}$	& $[T_\mathrm{d}]_\mathrm{peak}$	& $[T_\mathrm{d}]_\mathrm{mean}$ \\   
      & (degree)      &  (degree)     & (pc)     &   ({\Msolar})       &    ($\times 10^{22}$\thinspace cm$^{-2}$) & (K) & (K)\\
\hline 
\hline   						   	      	   						
1  &  188.951 &  0.882 &  0.39 &  498  & 36.48 & 26 & 19\\
2  &  189.030 &  0.781 &  0.35 &  274  & 22.30 & 36 & 21\\
3  &  189.032 &  0.808 &  0.32 &  261  & 12.69 & 22 & 18\\
4  &  189.032 &  0.790 &  0.25 &  128  & 7.10  & 27 & 20\\
5  &  189.014 &  0.826 &  0.24 &   91  & 3.72  & 17 & 15\\
6  &  189.007 &  0.839 &  0.25 &   74  & 3.66  & 17 & 15\\
7  &  188.975 &  0.911 &  0.19 &   46  & 3.21  & 15 & 14\\
8  &  188.966 &  0.922 &  0.37 &  156  & 3.19  & 15 & 13\\
9  &  188.991 &  0.871 &  0.30 &   97  & 3.03  & 16 & 15\\
10 &  188.992 &  0.859 &  0.21 &   51  & 2.70  & 16 & 15\\
11 &  188.977 &  0.904 &  0.26 &   70  & 2.63  & 16 & 15\\
12 &  188.953 &  0.903 &  0.30 &   105 & 2.55  & 24 & 17\\
13 &  188.910 &  0.897 &  0.32 &   101 & 2.53  & 17 & 14\\
14 &  188.974 &  0.918 &  0.30 &   106 & 2.43  & 15 & 13\\
15 &  188.958 &  0.889 &  0.22 &   54  & 2.43  & 26 & 20\\
16 &  188.902 &  0.762 &  0.16 &   27  & 1.83  & 14 & 14\\
\hline          
\end{tabular}
\end{table*}
\subsubsection{Study of {\it Herschel} clumps and their physical parameters}
\label{mas_clumps}
To study the fragmentation (or to detect clumps) in the column density map, we employed the {\it clumpfind} algorithm \citep[see for details][]{Williams_1994}. A total of 16 clumps are identified in the column density map, and their boundaries and IDs are shown in Figure~\ref{fig4}f. The mass ($M_\mathrm{clump}$) of each clump can be estimated using,
\begin{equation}
M_{\mathrm {clump}} = \mu_{\mathrm H_\mathrm{2}} m_{\mathrm H} \mathrm {Area}_{\mathrm {pixel}} \Sigma N(\mathrm H_\mathrm{2})
\end{equation}
Here, $m_{\mathrm H}$ is the mass of a hydrogen atom, $\mu_{\mathrm H_\mathrm{2}}$ is the mean molecular weight, $\mathrm {Area}_{\mathrm {pixel}}$ is the area subtended by one pixel, and $\Sigma N(\mathrm H_2)$ is the total column density. The value of $\mu_{\mathrm H_\mathrm{2}}$ is adopted to be 2.8. 
Various physical parameters (including mass) of these 16 clumps are listed in Table~\ref{tab1}. The clump masses range from 27 to 498\,$M_\odot$. Massive clumps (IDs: \#1 and \#2) are located toward the massive star-forming regions T1 and T2, respectively. Additionally, several fragments/clumps (IDs: \#3, \#4, \#5, \#6, \#9, and \#10; see also Table~\ref{tab1}) are also seen along their connecting filament F2 (see Figure~\ref{fig4}f). 
\subsubsection{The physical parameters of the sub-filaments}
\label{along_fil}
In order to understand the variation of the column density and dust temperature along the sub-filaments toward AFGL\,5180 and AFGL\,6366S, we have chosen several circular regions of radius 7{\rlap{\as}}~~~along the sub-filaments. In Figure~\ref{fig_SAVE}a, the {\it getsf} identified sub-filaments toward AFGL\,5180 and AFGL\,6366S are only shown with gray skeletons on the {\it Herschel} column density map. The selected circular regions along the sub-filaments are also overplotted. The average column density and dust temperature values for these circular regions are plotted as a function of distances along the sub-filaments from corresponding hubs in Figures~\ref{fig_SAVE}b and ~\ref{fig_SAVE}c, respectively. For both AFGL\,5180 and AFGL\,6366S, lengths of the associated filaments are within 1--2\,pc. All the sub-filaments toward AFGL\,6366S are traced with higher dust temperatures ({\simi}16--23\,K) and lower column density ({\simi}5$\times$10$^{21}$--2$\times$10$^{22}$\,cm$^{-2}$) compared to sub-filaments of AFGL\,5180 ($T_\mathrm d$ {\simi}13--19\,K and $N$(H$_2$) {\simi}5$\times$10$^{21}$--2$\times$10$^{22}$\,cm$^{-2}$). For sub-filaments, the average dust temperature and column density values peak at their junctions/hubs and then decrease along their lengths as moving away from the hubs. The trend of increasing column density along the filaments toward the hub has also been reported in one of the most well-known HFSs, Mon R2 \citep[see Figures 2 and 3 in][]{Kumar_2022MonR2}.
\subsection{Probing the distribution of molecular gas}
\label{sec:mol}
In this section, we have examined the CO line data to study the gas distribution in both the spatial and velocity space. 
The details of the molecular data are given in Section~{\ref{sec:obser}}. 
\subsubsection{Distribution and kinematics of molecular gas}
\label{sec:mol_mom}
In this section, we have presented the integrated intensity (moment-0) maps, the intensity-weighted velocity (moment-1) maps, and the intensity-weighted full width at half maximum (FWHM) linewidth map of the {\twco}, {\tco}, and {\cetno} emission in the direction of our target sites for an area, shown via a magenta rectangle in Figure~\ref{fig3}c. In all the moment maps, the NVSS radio continuum contour (at 3$\sigma$) tracing the boundary of the ionized region, the positions of the 6.7\,GHz MMEs, and the location of the massive O-type star are shown. In general, the {\cetno} emission is optically thin compared to the {\twco} and {\tco} emission, and traces the denser regions of the molecular cloud. All the moment-0 maps show almost identical morphology and support the existence of an elongated molecular cloud hosting T1 and T2 at its opposite edges. Both T1 and T2 show intense {\twco}/{\tco}/{\cetno} molecular gas emission compared to their filamentary connection. Figures~\ref{fig5}b,~\ref{fig5}e, and~\ref{fig5}h present the moment-1 maps of {\twco}, {\tco}, and {\cetno}, respectively. Noticeable velocity variations (of about 1-2\,{\kps}) are observed between star-forming sites (i.e., T1 and T2) and their connecting filament. The filaments detected from the dust continuum emission are overplotted on the moment-1 map for the {\tco} and {\cetno} emission (see Figures~\ref{fig5}e and ~\ref{fig5}h, respectively). The filamentary connection between two intense molecular emissions (at T1 and T2) is mainly detected in the velocity range [1.3, 2.7]\,{\kps} in Figures~\ref{fig5}e and ~\ref{fig5}h. 
Despite the coarse beam size of the molecular line data compared to {\it Herschel} 160\,{\micro} dust continuum image used to trace the filaments, the moment-0 maps show an elongated morphology spatially coexisting with filament F2. 
However, sub-filaments are not observed in the moment-0 maps. 
In Figures~\ref{fig5}c,~\ref{fig5}f, and~\ref{fig5}i, we display the linewidth maps of {\twco}, {\tco}, and {\cetno} data, respectively. Overall, {\twco} linewidth map shows the largest values, while {\cetno} presents the least in the same. But in all the lines, relatively higher line width values are found toward the edges of the filament (i.e., toward T1 and T2) compared to their connecting filament.

We have highlighted some circular regions of radius 30{\as} along the filament F2 in Figure~\ref{fig4}e, and various physical parameters along the filament are extracted. The average column density and dust temperature values for the circular regions are presented in Figures~\ref{fig6}a and~\ref{fig6}b, respectively. These values are estimated using the {\it Herschel} column density and  dust temperature maps, which are shown in Figures~\ref{fig4}c and~\ref{fig4}d, respectively. Higher values of column density and dust temperature are observed toward T1 and T2. Using the {\tco} and {\cetno} molecular line data, we have also computed average velocity, thermal sound speed ($a_{\mathrm s}$ = ($k T_{\mathrm {kin}}/\mu m_{\mathrm H})^{1/2}$), non-thermal velocity dispersion ($\sigma_{\mathrm {NT}}$), and the ratio of thermal to non-thermal pressure ($P_{\mathrm {TNT}} = {a_{\mathrm s}^2}/{\sigma^2_{\mathrm {NT}}}$) for the same circular regions. Here, $T_{\mathrm {kin}}$ is the kinetic gas temperature. For the calculations of different physical parameters for each circular region, we have assumed $T_{\mathrm {kin}}$ to be equal to the average dust temperatures.
The mean molecular weight per particle $\mu$ is adopted to be 2.37. 
The thermal and non-thermal velocity dispersion (i.e., $\sigma_{\mathrm {T}}$ and $\sigma_{\mathrm {NT}}$, respectively) for a molecular line can be derived using,
\begin{equation}
\sigma_{\rm T}$ = $(k T_{\mathrm {kin}}/{\chi} m_{\mathrm H})^{1/2},
\label{sigmathermal}
\end{equation}
\begin{equation}
\sigma_{\mathrm {NT}} = \sqrt{\frac{\Delta V^2}{8\ln 2}-\frac{k T_{\mathrm {kin}}}{\chi m_{\mathrm H}}} = \sqrt{\frac{\Delta V^2}{8\ln 2}-\sigma_{\mathrm T}^{2}} ,
\label{sigmanonthermal}
\end{equation}
where, $\chi$ is the mass of the molecules in terms of hydrogen atomic mass and $\Delta V$ is the measured linewidth for the molecular line data. Therefore, $\chi$ {\simi}29 and {\simi}30 for {\tco} and {\cetno}, respectively. From the average velocity and non-thermal velocity dispersion profile for {\tco} and {\cetno} as shown in Figures~\ref{fig6}c and \ref{fig6}d, respectively, it can be observed that T1 and T2 correspond to higher velocity and non-thermal velocity dispersion. Additionally, T1 shows  higher velocity than T2, and an oscillation in velocity profile is evident in both {\tco} and {\cetno} data. The profile for Mach number (i. e., $M = \frac{\sigma_{\mathrm {NT}}}{a_{\mathrm s}}$) and thermal to non-thermal pressure ratio are shown in Figures~\ref{fig6}e and~\ref{fig6}f, respectively. From the Mach number profile, derived using {\cetno}, we can find that the star-forming sites T1 and T2 are supersonic (with $M > 3$). However, Mach number profile for the {\cetno} data drops down to {\simi}2 toward the filamentary connection between T1 and T2 (see Figure~\ref{fig6}e). Mach number profile extracted using the {\tco} line data reveals the supersonic nature of the filament (i.e., $M >3$; see Figure~\ref{fig6}e). The filamentary structure F2 is traced with lower non-thermal velocity dispersion, leading to higher thermal to non-thermal ratios compared to T1 and T2 in {\cetno} data. In Figure~\ref{fig6}f, the values of $P_{\mathrm {TNT}}$ are different in the {\tco} and {\cetno} emissions. Because the {\cetno} data is optically thin compared to the {\tco} emission. Therefore the {\cetno} emission is less affected by the opacity broadening \citep{phillips_1979,hacar_2016,Yang2022}.  
\subsubsection{Identification of the molecular cloud components}
\label{sec:mol_1}
In order to probe the complete picture of the molecular gas distribution toward our target sites, we analyzed {\tco} data for an extended area of {\simi}1{\rlap{$^\circ$}}\,\,\,$\times$~1{\rlap{$^\circ$}}~~(or {\simi}26\,pc $\times$ 26\,pc). The extended moment-0 map of {\tco} data is shown in Figure~\ref{fig7}a for a velocity integration range of [-5.1, 15.1]\,{\kps}.
The moment-0 map presents a complex morphology of the molecular cloud.
Figure~\ref{fig7}b presents {\it Hersechel's} view of the extended region at 250\,{\micro}. The {\it Herschel} dust continuum data shows the elongated filamentary structure embedded in the molecular cloud, hosting the 6.7\,GHz MMEs at its opposite edges. In Figure~\ref{fig7}c, the moment-1 map for the {\tco} data infers intensity-weighted velocity distribution. Two extreme velocity ranges can be observed in the moment-1 map based on the colorbar in colors blue and red. The average intensity profile extracted for the rectangular region (in black) presented in Figure~\ref{fig7}c is shown in the inset. It shows two velocity components with their peaks at the velocities {\simi}2.7\,{\kps} and {\simi}6.4\,{\kps} for the blue-shifted and red-shifted components, respectively (see the blue and red vertical lines). The blue-shifted and red-shifted components are separated by a velocity of about 3.7\,{\kps}. 
To determine the velocity ranges for the cloud components, the Galactic latitude--velocity ({\it b}--{\it v}) diagram is presented in Figure~\ref{fig7}d. The {\it b}--{\it v} diagram is presented for the area shown by white dotted rectangle in Figure~\ref{fig7}a. This diagram reveals a clear signature of the presence of two cloud components; the blue-shifted (hereafter, blue) component: [$-$3.1, 4.8]\,{\kps} and the red-shifted (hereafter, red) component: [5.8, 12.9]\,{\kps}. Both the cloud components are found to be connected with weak molecular emission in an intermediate velocity range [4.8, 5.8]\,{\kps}, which is shown with two yellow dashed lines in Figure~\ref{fig7}d.
\subsubsection{Spatial distribution of molecular cloud components}
\label{sec:mol_2}
The moment-0 maps for the blue and the red cloud components are shown in colors cyan and red, respectively, in Figure~\ref{fig8}a. The moment-0 map of the red component is only displayed within the boundaries of the yellow dashed rectangle. A complementary spatial distribution of the cloud components can be observed in Figure~\ref {fig8}a. However, a more convincing picture of their complementarity is achieved by shifting the red cloud component toward the south-west direction by about 2.3\,pc, presented in Figure~\ref{fig8}b. The initial (i.e., before displacement) and the final (i.e., after displacement) positions of the red cloud component are marked by a dashed and a solid yellow rectangle, respectively. 
The massive star-forming sites (i.e., the position of the 6.7\,GHz MMEs indicated with the triangles) are spatially found toward the junction of two cloud components (see in Figure~\ref{fig8}b).\\ 

Section~\ref{sec:dis} presents the implication of these observed results.
\section{Discussion}
\label{sec:dis}
Using the {\it Herschel} image at 160\,{\micro}, the outcome of the {\it getsf} utility shows two interesting structures in our target area: the curved feature F1 and the elongated filament F2. The curved feature F1 is associated with the PAH emission at 3.3\,$\mu$m and is located at the boundary of the {\htwo} region, S247. The edges of filament F2 are located within sites T1 and T2, both of which also host HFS. Signatures of outflow activities and YSOs are found toward the entire filament F2. Interestingly, the groups of YSOs and the 6.7\,GHz MMEs are mainly concentrated at the opposite edges of F2. Using the {\twco}, {\tco}, and {\cetno} emission, both the HFSs are depicted with supersonic and non-thermal gas motions having higher Mach number and lower thermal to non-thermal pressure ratio (see Figure~\ref{fig6}). The velocity dispersion is found to be higher at the filament's edges compared to the central region, suggesting intense star formation activities at the edges. Earlier studies proposed the CCC and triggered star formation scenarios in our selected target area. Based on our observational findings, we have carefully assessed these scenarios and the role of filaments in the below sections.
\subsection{Impact of the massive star on its environment}
\label{sec:dist2}
In general, Lyman continuum photons of a massive star ionize the surrounding medium and create an {\htwo} region. This {\htwo} region expands supersonically in the surrounding neutral gas medium, which drives a shock. Based on this property, in the literature, the ``collect and collapse'' \citep[C\&C;][]{Elmegreen_1977} and the radiatively driven implosion \citep[RDI;][]{bertoldi89} scenarios have been proposed to explain triggered star formation. In the RDI model, star formation activities are produced by the compression of the pre-existing dense clumps by the shock wave. In the C\&C scenario, the shock continues to sweep up cool neutral gas to form a massive and dense shell around the {\htwo} region \citep{Deharveng_2005,dale_2007}. Over time this shell of collected gas becomes gravitationally unstable, and collapses to form new stars. 

In our target area, a cavity in the molecular gas distribution is seen in all the moment-0 maps (see the cyan contour in Figures~\ref{fig7}a,~\ref{fig7}d, and~\ref{fig7}g). This cavity in molecular gas shows a footprint of the interaction between the molecular gas and ionizing photons. The curved feature F1, which is deprived of molecular emission, is located close to the massive O-type star (i.e., cgo115). In contrast, filament F2 traced in the molecular and dust continuum maps appears further away from F1 and cgo115. Considering an excess of PAH 3.3\,$\mu$m emission in the {\it Spitzer} 3.6\,$\mu$m band, the impact of the O-type star on the curve feature F1 is likely. 
All the derived pressure components (i.e., P$_{\mathrm{wind}}$, P$_{\mathrm{rad}}$, and P$_{\Hii}$) have consistent values along F1. Among them, P$_{\Hii}$ is the most dominating while P$_{\mathrm{wind}}$ contributes the least. The combined pressure value along F1 is about 10$^{-10}$\,dynes\,cm$^{-2}$, well above the typical molecular gas pressure \citep[$\sim$10$^{-11}$\,dynes~cm$^{-2}$; see Table 7.3 in][]{dyson97}. 
Therefore, the feedback from cgo115 seems sufficient to form the feature F1 and may be responsible for its curved appearance. 
Similar pressure analysis along the filament F2 (see Figure~\ref{fig2}c) shows that every pressure component falls as we move away from the central point. Interestingly, as the {\htwo} region is not directly associated with F2 (see the cyan contour in Figure~\ref{fig2}a), therefore, we can exclude P$_{\Hii}$ while computing combined pressure. Now, the total of wind and radiation pressure along F2 peaks at around 1.5 $\times$ 10$^{-11}$\,dynes\,cm$^{-2}$; hence it can be balanced by the molecular gas pressure. Including P$_{\Hii}$, we find the peak of the combined pressure around 10$^{-10}$\,dynes\,cm$^{-2}$. Therefore, cgo115 has the potential to impact F2 in the future. But the current pressure values along F2 and its linear morphology do not allow us to connect its origin due to the feedback of the massive star, cgo115. The combined pressure values of cgo115 at the massive star-forming sites T1 and T2 (residing at the edges of F2) are also not significant to affect the star-formation activity. Therefore, we do not find triggered star formation scenarios feasible in these star-forming regions. This proposal agrees with the previously reported work of \citet{Vasyunina_2010}. By analyzing NIR spectroscopic data of the AFGL\,5180 cluster, \citet{Vasyunina_2010} found that the more evolved stars are located further away from the {\htwo} region, whereas the younger sources that are driving outflows are situated closer to the {\htwo} region. This signature is opposite to the prediction of the triggered star formation scenarios \citep{dale_2007,Dale_2015}.
\subsection{Role of filaments toward our target sites}
\label{sec:role_fil}
A small cluster of the radio continuum sources toward the site T1 and the 6.7\,GHz MMEs toward the sites T1 and T2 together support the ongoing MSF activities in both the HFSs. Junctions of the sub-filaments (i.e., hubs) show higher column density and dust temperature toward T1 and T2. We suggest that the mass accumulation through sub-filaments may be the possible scenario for the MSF activity toward T1 and T2 \citep[e.g.,][]{myers_2009,tan14,tige_2017,motte_2018,trevino19,rosen_2020,Dewangan2022}. Concerning the observed hub-filament configurations, the proposed paradigm of MSF by \citet{motte_2018} starts with the formation of density-enhanced hubs at the junctions of several filaments. These hubs host massive dense cores (MDCs), which harbor low-mass prestellar cores during their starless phase. Over time these MDCs become protostellar in a time scale of around 3 $\times$ 10$^{5}$\,yrs. They host infra-red (IR) quite low-mass stellar embryos ($<$ 8\,{\Msolar}) at this stage. They grow into IR bright high-mass ($>$ 8\,{\Msolar}) protostars via gravitationally driven inflows, and then these sources become massive stars that excite {\htwo} regions \citep[see more details in][]{motte_2018}. This scenario supports the picture of a mass accumulation from the large-scale to the dense core-scale through filaments \citep{tige_2017,motte_2018,trevino19}. Earlier, \citet{minier_2001} suggested that the 6.7\,GHz MMEs most likely trace the massive protostellar phase in the early stages of MSF. Concerning the evolutionary paradigm of MSF, \citet{motte_2018} proposed that the massive protostar is equivalent to the ``high-mass protostellar phase.'' 

A recent study by \citet{Kumar_2020} aimed to identify HFSs in the Milky Way. They analyzed approximately 35000 Hi-GAL clumps located within the Galactic longitude range of 68\rlap{$^\circ$}~~$\geq l \geq$ -72\rlap{$^\circ$} ~and a Galactic latitude range of |b| $\leq$ 1\rlap{$^\circ$}~~~from the study of \citet{Elia_2017}. To identify the filamentary skeletons associated with the clumps, they used the DisPerSE algorithm \citep{Sousbie_2011} on 10{$\rlap{\am}$} ~$\times$ 10${\rlap{\am}}$ ~cut-outs of Herschel 250\,{\micro} images. The study found around 3700 candidate HFSs, constituting approximately 11\% of the total clumps analyzed. The filaments forming the HFSs had a mean length of about 10-20\,pc. The study also found that all the clumps with luminosity greater than 10$^4$ and 10$^5$\,L${\odot}$, at distances within 2\,kpc and 5\,kpc, respectively, were located in the hubs of HFSs. Typically, the hubs were found to have 3-7 skeletons joining at the junction. In this context, the identification of HFSs in AFGL\,5180, and AFGL\,6366S provides further insights into the nature and properties of these structures. The Hi-GAL clumps toward AFGL\,5180 and AFGL\,6366S have luminosities of {\simi}10$^3$ and {\simi}10$^4$\,L${\odot}$ \citep{Elia_2021}, respectively. Both the clumps host 6.7\,GHz MMEs as the signpost of massive star-forming activity, and they are associated with at least 4 filaments detected based on {\it Herschel} 160\,{\micron} image. From the diameter of the circular regions (i.e., 14{\rlap{\as}}~~~) used to determine the average properties along the filaments, the widths of the detected filaments can be roughly estimated to be {\simi}0.1 pc, which is consistent with the typical width of Herschel filaments \citep{Andre_2022}. The column density and dust temperature values increase along the filaments in the direction to the hubs. The flow of dust and gas through the filaments amplifies the column density of the hub, suiting it for MSF \citep{myers_2009,Kumar_2020,Dewangan2022}. The length of the filaments participating in the formation of HFSs toward our target sites is nearly 1--2\,pc. Therefore, for nearby regions ($d\lesssim$ 2\,kpc), one can expect more HFSs with parsec-scale filaments. This proposal also agrees with the ﬁlament selection criteria of \citet{Kumar_2020} because the length scale associated with the image's resolution increases with distance. It is also noticeable that, though the Hi-GAL clump at AFGL\,5180 has a luminosity of {\simi}10$^3$\,L${\odot}$, it is an observed massive star-forming site. Therefore, we can also expect more such sources in the analysis of 360{$^\circ$}~~{\it Herschel} data. Hence, a similar analysis like \citet{Kumar_2020} for the Hi-GAL 360\rlap{$^\circ$}~~ clumps \citep{Elia_2021} with {\it Herschel} 160/250\,{\micro} images can reveal more details about HFSs.    

In star formation study, it is evident that molecular clouds often create filamentary structures, which further fragment into dense cores \citep[e.g.,][]{andre_2010}. Our molecular gas study shows a similar filamentary structure further fragmented in dense cores toward its edges. Recently, \citet{Clarke_2020} found that the filaments are more prone to fragment into sub-filaments, which are responsible for the formation of hubs via their merging event, and the cores formed at the filament's edges are more massive than the interior ones. 
The coarse beam sizes of the molecular line data are one of the key obstacles in studying such aspects. Therefore, such a study is beyond the scope of this work.

A connection between CCC and the origin of filaments and HFSs was found in Smoothed particle hydrodynamics simulation by \citet{balfour15} for head-on colliding clouds. According to \citet{balfour15}, CCC produces a shock-compressed layer, which fragments into the filaments. For lower collision velocity, filaments predominantly appear radial, while for higher collision velocity formation of filaments looks like a spider's web. Recent works of \citet{fukui19ex}, \citet{tokuda19ex}, and \citet{maity_W31} presented the observational connections between CCC and the formation of HFS.
\subsection{Cloud-cloud collision scenario}
\label{sec:CCC}
The CCC scenario is an effective triggering mechanism of MSF. Molecular clouds often collide due to intra-molecular cloud velocity dispersion \citep[typically with a collision rate of 1 in every 100 years;][]{tasker_2009,dobbs_2015}. If the collision speed is supersonic, it creates a shock-compressed interface with an increased effective sound speed \citep{habe92,anathpindika_2010,fukui_2014,fukuia_2018}. It provides a higher mass accretion rate leading to higher ram pressure to overcome the radiation pressure. The observational signatures of CCC are the bridge feature in the position-velocity (PV) space and the complementary spatial distribution of the cloud components \citep[e.g.,][]{torri_2011,torri_2015,torri_2017,torri_2021,fukui_2014,fukui_2015,CCC3,fukuia_2018,fukui21,dewangan_2017,nishimura_2017,hayashi_2018,sano_2017,sano_2018,fujita21}. The bridge feature is a low-intensity connection between two cloud components in the PV space \citep[e.g.,][]{haworth_2015,torri_2017,Dewangan2017,Priestley2021}. The collision of clouds results in a cavity in the larger cloud due to the conservation of linear momentum, referred to as the complementary distribution. 

Based on the analysis of $^{12}$CO(J = 2-1) data (beam size {\simi}2{\rlap{\am}.7}), earlier \citet{Shimoikura_2013} proposed CCC as a possible triggering mechanism in our target sites (see Figures 3b and 4 in their paper). They reported two clouds at the velocities of [$-$3, 5] and [5, 13]\,{\kps}. In Section~\ref{sec:mol_1}, we have also found the existence of two clouds (around [$-$3.1, 4.8] and [5.8, 12.9]\,{\kps}), which are connected by a bridge feature in velocity space. 
Additionally, we have investigated a spatial ﬁt of ``key/intensity-enhancement'' and ``cavity/keyhole/intensity-depression'' features (i.e., complementary distribution) of two clouds, which is further strengthened after a spatial shift of about 2.3\,pc in the red cloud component.
Previously, \citet{fukuia_2018} detected an identical signature of CCC (i.e., complementary distribution with a spatial shift)  toward M43 for the formation of massive stars in the Orion Nebula Cluster \citep[see Figures 10 and 11 in][]{fukuia_2018}. The PV diagrams depend upon the angle ($\alpha$) between the axis of collision and the line of observation, which is thoroughly explained in \citet{fukuia_2018}. Earlier, \citet{maity_W31} assumed $\alpha${\simi}10{\rlap{$^\circ$}}\,\, based on minute skewness of the ``V'' like feature in PV diagram for the CCC scenario in the W31 complex. It is important to note that the PV diagram also depends upon the spatial morphology of the molecular cloud components. Therefore, the detection of two cloud components toward our target sites in the PV diagram ensures that the angle ($\alpha$) is below 90{\rlap{$^\circ$}} ~~\citep{fukuia_2018,fukui_2021}. On the other hand, the requirement of spatial shift for the complementary distribution hints the angle $\alpha$ to be greater than 0{\rlap{$^\circ$}} ~~\citep{fukuia_2018,fukui_2021}. Therefore, it is logical to use the $\alpha$ value in between 30{\rlap{$^\circ$}} ~~to 60{\rlap{$^\circ$}} ~~ for the estimation of the collision time scale. If the observed velocity difference between the colliding clouds is $V_{\mathrm{obs}}$, then along the collision axis it would be, $V_{\mathrm{loc}} = \frac{V_{\mathrm{obs}}}{\cos\alpha}$. Similarly, spatial shift along the collision axis would be, $l_{\mathrm{loc}} = \frac{l_{\mathrm{obs}}}{\sin\alpha}$, for the observed spatial shift of $l_{\mathrm{obs}}$ in the sky plane. Finally, the collision time scale can be obtained by, $t_{\mathrm{collision}} = \frac{l_{\mathrm{loc}}}{V_{\mathrm{loc}}}$. In case of present study, the observed velocity and spatial shift are $V_{\mathrm{obs}} = 3.7$\,{\kps} and $l_{\mathrm{obs}} = 2.3$\,pc, respectively. Thus for various values of $\alpha$, i.e., $\alpha$ = 30$^\circ$, 45$^\circ$, and 60$^\circ$, $t_{\mathrm{collision}}$ is found to be 1.05, 0.60, and 0.35\,Myr, respectively. Hence, a collision between molecular clouds about 1\,Myr ago initiated the massive star-forming activity toward our target sites. However, it is interesting to note that previous studies determined the age of the NIR cluster to be {\simi}2.5\,Myr \citep{devine_2008,Vasyunina_2010}. This infers that the colliding clouds were active in low-mass star formation well before the collision took place. As earlier mentioned, CCC is emerging as a potential mechanism of HFS formation in both theoretical and observational studies \citep{balfour15,fukui19ex,tokuda19ex,maity_W31}. It is also interesting to note that our target sites host two HFSs at the edges of a primary filament F2; to date, only a few such systems are reported in the literature (e.g., the filamentary clouds G45.1+0.3 \citep{bhadari22}, IC\,5146 dark Streamer \citep{Wang_2019,Dewangan_2023}, and NGC 6334 \citep{Zernickel_2013,Arzoumanian_2021}. These results thus open a new window for discussion for the formation of HFSs at the edges of a filament.
\subsection{End-dominated collapse scenario}
\label{sec:EDC}
\citet{Wang_2019} proposed a scenario for the formation of HFSs at the edges of a main filament in their study of IC\,5146 dark Streamer, which includes the role of the magnetic field in addition to the gravitational instability of an isolated filament. Once a primary filament becomes supercritical (both thermally and magnetically), it collapses along its major axis because of gravity. Due to the effect of end-dominated collapse (EDC) or edge collapse, massive fragments are exclusively found at the edges of the filament. According to \citet{Bestien_1983,pon_2012,Clarke_2015}, the differential gravitational acceleration along the longer axis of an isolated filament drives EDC depending upon the aspect ratio of the filament. During this process, the ram pressure exerted by gas motion is sufficient to pinch the magnetic field lines and forms the U-shaped magnetic field \citep[or bending effect; see more details in][]{Gomez_2018,Wang_2019} at the filament's edges. A recent study by \citet{Wang_2019} and \citet{Chung_2022} suggested that the EDC filament IC\,5146 prefers to show the bending effect at it's egdes. Using {\it Planck} polarization data, \citet{Dewangan_2023} verified this bending effect in nearby ($d\lesssim$ 2\,kpc) EDC filaments (i.e., NGC 6334; \citet{Zernickel_2013}, S242; \citet{dewangan_2019,yuan_2020}, IC\,5146; \citet{Wang_2019,Chung_2022}, and Mon R1; \citet{Bhadari_2020}). Finally, the massive components accumulated at the edges of the primary filament fragments further along the curved magnetic fields to form HFSs \citep[see Figure~13 in][]{Wang_2019}. 

In literature, apart from IC\,5146, the filamentary clouds G45.1+0.3 \citep{bhadari22} and NGC\,6334 \citep{Zernickel_2013,Arzoumanian_2021} serve as the sites where the EDC and HFSs are simultaneously investigated. Although signatures of MSF, massive dust clumps, higher column densities, and clusters of YSOs are mainly depicted toward both the edges of the filament F2, our analysis of {\it Planck} polarization data (see the details in Appendix~\ref{sec:posbfield}) does not clearly show the bending effect (see Figure~\ref{fig9}a) to draw a conclusive statement on the EDC process. New high-resolution polarization data toward our target area can shed more light on the bending effect toward both edges of F2. 
In addition, we conducted a core-scale study of the AFGL\,5180 region to investigate the outflow activity and physical association between the 6.7\,GHz MME and NVAS 8.46\,GHz radio continuum emission with embedded dust cores. To accomplish this, we utilized high-resolution NIR images and ALMA 1.3\,mm dust continuum emission. The results of our study are presented in Appendix~\ref{sec:ALMA_res}.    

Altogether our observational results suggest that CCC had initiated the star-forming activity around 1\,Myr ago toward the massive star-forming sites AFGL\,5180 and AFGL\,6366S. 
Also, the observed HFSs at both edges of the filament can be the consequence of the CCC scenario. However, we suggest that the magnetic field and differential gravitational acceleration can also shape the observed morphology of the target sites. 
In the context of mass accumulation toward our target sites, we suspect that the clumps in the central hub accumulate materials through filamentary accretion. Then it is also possible that individual cores may grow in mass, sharing the common source of gas and dust within the clump. In other words, the final masses of the stars inside the clump are determined not only by the small clump- or core-scale mass accretion but also by the larger scale, filamentary mass accretion.
\section{Summary and Conclusions}
\label{sec:conc}
To unveil star formation processes, we have performed a multi-scale and multi-wavelength study of two massive star-forming sites, AFGL\,5180 and AFGL\,6366S. This work includes a careful analysis of various data sets from NIR, mid-infrared, sub-millimeter, and centimeter wavelengths. It also includes the analysis of different molecular line data (i.e., {\twco}, {\tco}, and {\cetno}).
The major observational outcomes of this work are as follows:
\begin{enumerate}
\item AFGL\,5180 and AFGL\,6366S are nearby (distance {\simi}1.5\,kpc) massive star-forming sites in their earlier stages. Both the sites host a Class\,{\sc II} 6.7\,GHz MME and reside at the opposite edges of an elongated filament F2, traced in dust continuum emission at the periphery of the {\htwo} region, S247. 
\item Application of the {\it getsf} utility on the {\it Herschel} 160\,{\micron} image reveals a HFS toward both the sites, hosting at least one 6.7\,GHz MME. The radio continuum map at 8.46\,GHz reveals a small cluster of radio sources in the vicinity of the central hub of the HFS toward AFGL\,5180.

\item Based on the analysis of the photometric data at 1--5\,$\mu$m, a total of 208\,YSOs are identified toward our target area. The YSOs and outflow signatures are traced along the filament F2, which is further supported by the {\it Spitzer} ratio map. Clusters of YSOs are primarily depicted toward the edges of the filamentary cloud.

\item The filamentary cloud seems spatially close to the {\htwo} region, S247 excited by a massive O9.5 star, cgo115. 
Various pressure components exerted by the O-type star (i.e., P$_{\Hii}$, P$_{\mathrm{rad}}$, and P$_{\mathrm{wind}}$) on its surroundings are estimated. In this connection, the impact of the energetic feedback from the massive star on the filamentary cloud is found to be insignificant.

\item Based on the analysis of molecular line data, previously proposed CCC scenario seems to be applicable in our target sites. 
\end{enumerate}
Overall, the collision of two clouds at [$-$3.1, 4.8]\,{\kps} and [5.8, 12.9]\,{\kps} had occurred about 1\,Myr ago. Hence, the CCC process appears to explain the observed star formation activities (including massive stars) and HSFs. The filament connecting to sites AFGL\,5180 and AFGL\,6366S, can be a candidate of EDC. To further confirm the EDC process, high-resolution polarimetric and spectroscopic observations in sub-millimeter wavelengths will be helpful.

\section*{Acknowledgments}
We thank the anonymous referee for providing the valuable comments and suggestions, that improved the scientific content of this paper. The research work at Physical Research Laboratory is funded by the Department of Space, Government of India. We acknowledge F. Navarete for providing us with the narrow-band H$_2$ and K-band continuum images. This work is based [in part] on observations made with the {\it Spitzer} Space Telescope, which is operated by the Jet Propulsion Laboratory, California Institute of Technology, under a contract with NASA. This research made use of the data from the Milky Way Imaging Scroll Painting (MWISP) project, which is a multi-line survey in $^{12}$CO/$^{13}$CO/C$^{18}$O along the northern galactic plane with PMO-13.7m telescope. We are grateful to all the members of the MWISP working group, particularly the staff members at the PMO-13.7~m telescope, for their long-term support. MWISP was sponsored by National Key
R\&D Program of China with grant 2017YFA0402701 and by CAS Key Research Program of Frontier Sciences with grant QYZDJ-SSW-SLH047. This research has made use of the NASA/IPAC Infrared Science Archive, which is funded by the National Aeronautics and Space Administration and operated by the California Institute of Technology. The NVAS image was produced as part of the NRAO VLA Archive Survey, (c) AUI/NRAO. This research is based [in part] on observations made with the NASA/ESA Hubble Space Telescope obtained from the Space Telescope Science Institute, which is operated by the Association of Universities for Research in Astronomy, Inc., under NASA contract NAS 5–26555. These observations are associated with the proposal id 14465. This paper makes use of the following ALMA data: ADS/JAO.ALMA\#2015.1.01454.S. ALMA is a partnership of ESO (representing its member states), NSF (USA) and NINS (Japan), together with NRC (Canada), MOST and ASIAA (Taiwan), and KASI (Republic of Korea), in cooperation with the Republic of Chile. The Joint ALMA Observatory is operated by ESO, AUI/NRAO and NAOJ. This research made use of {\it Astropy}\footnote[1]{http://www.astropy.org}, a community-developed core Python package for Astronomy \citep{astropy13,astropy18}. For figures, we have used {\it matplotlib} \citep{Hunter_2007} and IDL software.  
\subsection*{Data availability}
The {\it Herschel} and {\it Spitzer} data underlying this work are available in the publicly accessible NASA/IPAC infrared science archive\footnote[2]{https://irsa.ipac.caltech.edu/frontpage/}. The K-band image is available in the UKIDSS-GPS survey\footnote[3]{http://wsa.roe.ac.uk/.}. The NVSS radio continuum image is available\footnote[4]{https://skyview.gsfc.nasa.gov/current/cgi/query.pl}. ALMA fits files utilized in this work are accessible from ALMA Data Archive\footnote[5]{http://jvo.nao.ac.jp/portal/alma/archive.do}. {\it Gaia} EDR3 can be obtained from the publicly accessible server\footnote[6]{ https://gea.esac.esa.int/archive/}. Distance of the {\it Gaia} EDR3 sources is available in the server\footnote[7]{https://vizier.cds.unistra.fr/viz-bin/VizieR?-source=I/352}. HST NIR images are available in HST Data Archive\footnote[8]{https://www.cadc-ccda.hia-iha.nrc-cnrc.gc.ca}. NVAS radio continuum can be obtained from NRAO Data Archive\footnote[9]{https://science.nrao.edu}. {\it Planck} dust polarization data utilized in this work can be found in the publicly accessible server\footnote[10]{https://irsa.ipac.caltech.edu/applications/{\it Planck}/}.

%
\begin{figure*}
\includegraphics[width= 15 cm]{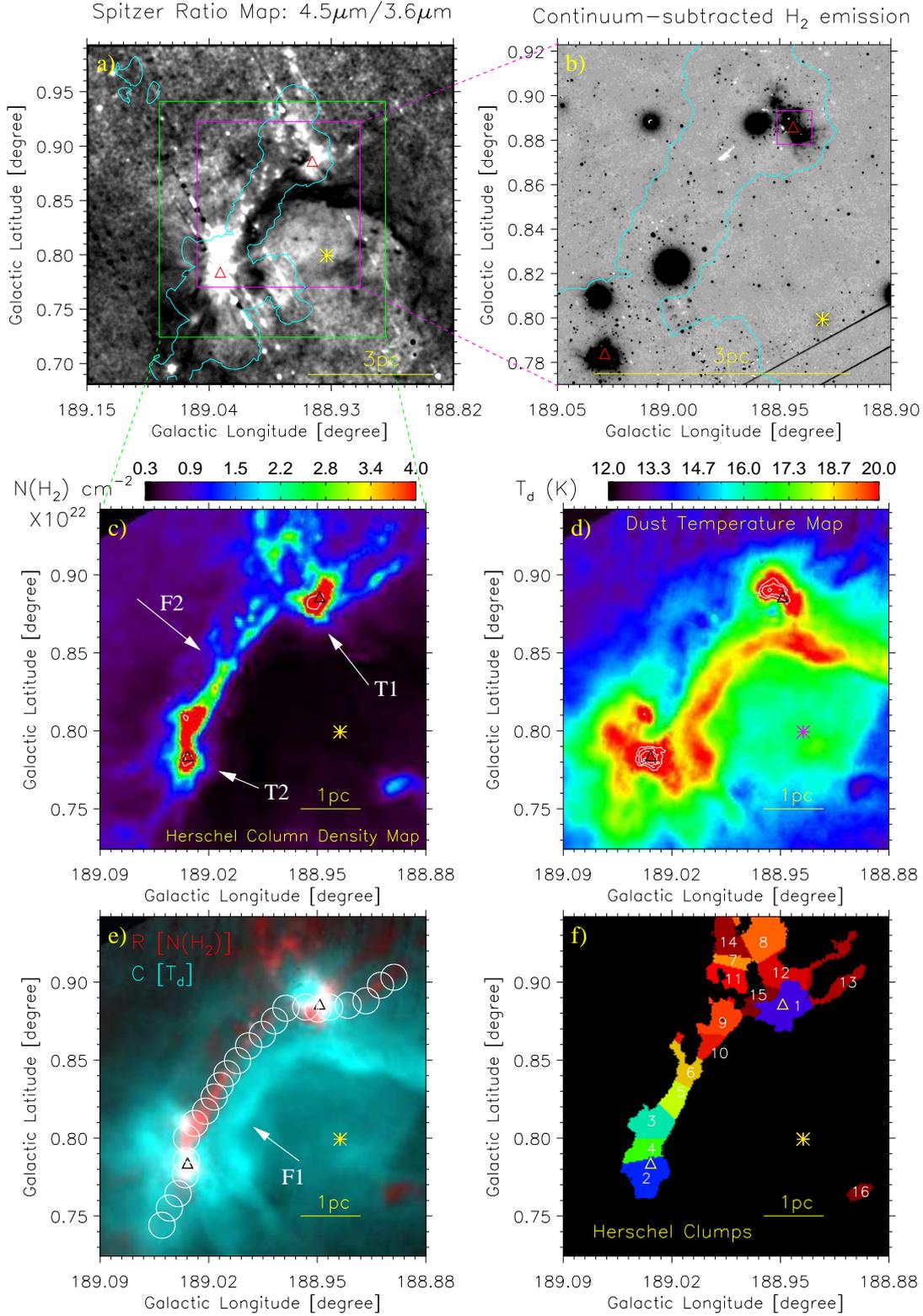}
\caption{a) Overlay of the YSOs surface density contour at 15\,YSOs~pc$^{-2}$ on the {\it Spitzer} ratio map (i.e., 4.5\,$\mu$m/3.6\,$\mu$m). b) The panel shows the continuum-subtracted H$_2$ map (at 2.12\,$\mu$m) in grayscale for an area indicated by a magenta rectangle in Figure~\ref{fig4}a. The magenta rectangle overplotted toward AFGL\,5180 presents the area utilized for the zoomed-in view in Figure~\ref{fig10}. c) The {\it Herschel} column density map (for an area indicated by a green rectangle in Figure~\ref{fig4}a), overplotted with a white contour at the column density value of 10$^{23}$\,N(H$_2$) cm$^{-2}$. d) The {\it Herschel} dust temperature map (for the same area as Figure~\ref{fig4}c), overlaid with contours (in white) at the levels of 22, 24, and 26\,K. e) A two-color composite image using the {\it Herschel} column density (in red) and dust temperature map (in cyan). Several circles (of radius 30{\as}) are also marked in the panel, where the average values of various physical parameters are computed (see Figure~\ref{fig6} and Section~\ref{sec:mol_mom} for details). f) The panel shows the boundaries of the clumps identified in the {\it Herschel} column density map using the IDL-based algorithm {\it clumpfind}. The last four panels show a scale bar of 1\,pc. Other symbols are identical to Figure~\ref{fig1}a.}
\label{fig4}
\end{figure*}
\begin{figure}
\includegraphics[width=0.5\textwidth]{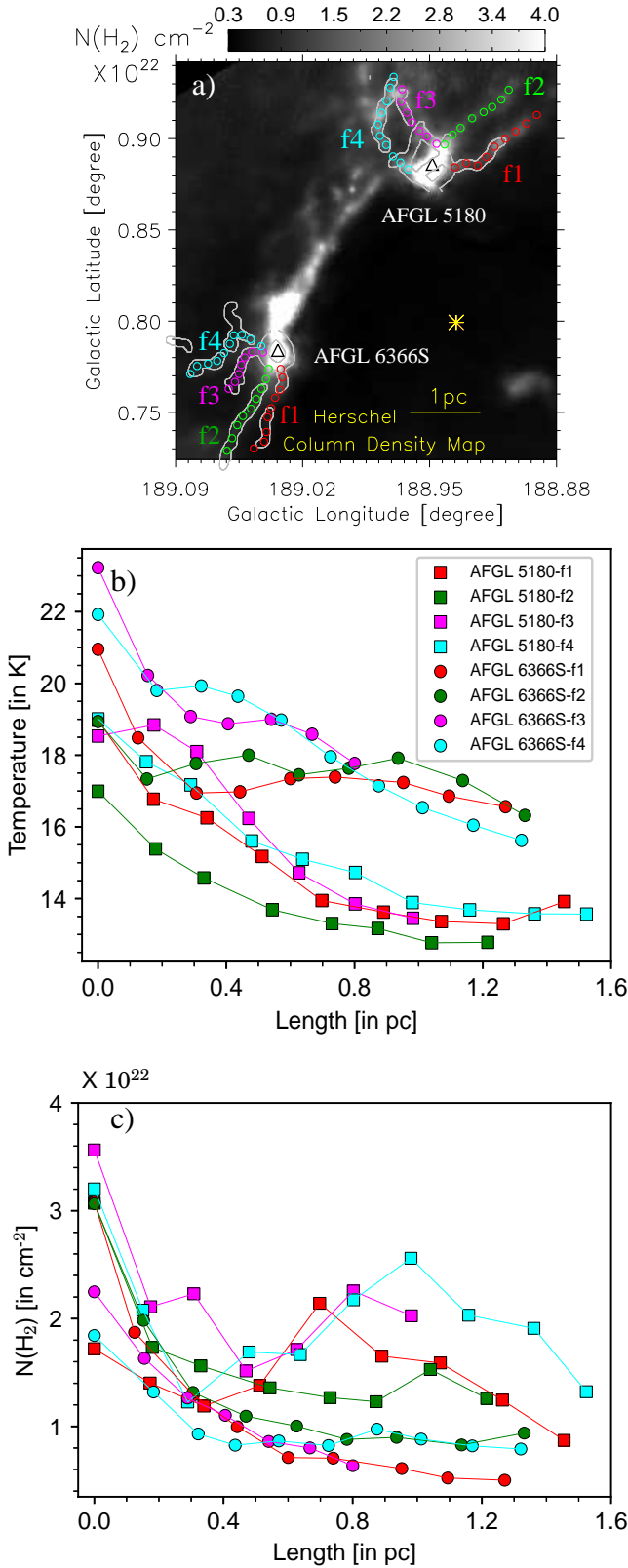}
\caption{ a) {\it Herschel} column density map identical as Figure~\ref{fig4}c in grayscale. The sub-filaments (e.g., f1, f2, f3, and f4) both toward AFGL\,5180 and AFGL\,6366S are shown with gray contours. Other symbols are the same as Figure~\ref{fig4}c. b) Distribution of the average dust temperature along the sub-filaments for the circular regions (of radius 7\rlap{\as}~) overplotted on panel ``a''. c) Distribution of the average column density along the sub-filaments for the same circular regions.}
\label{fig_SAVE}
\end{figure}

\begin{figure*}
\includegraphics[width= \textwidth]{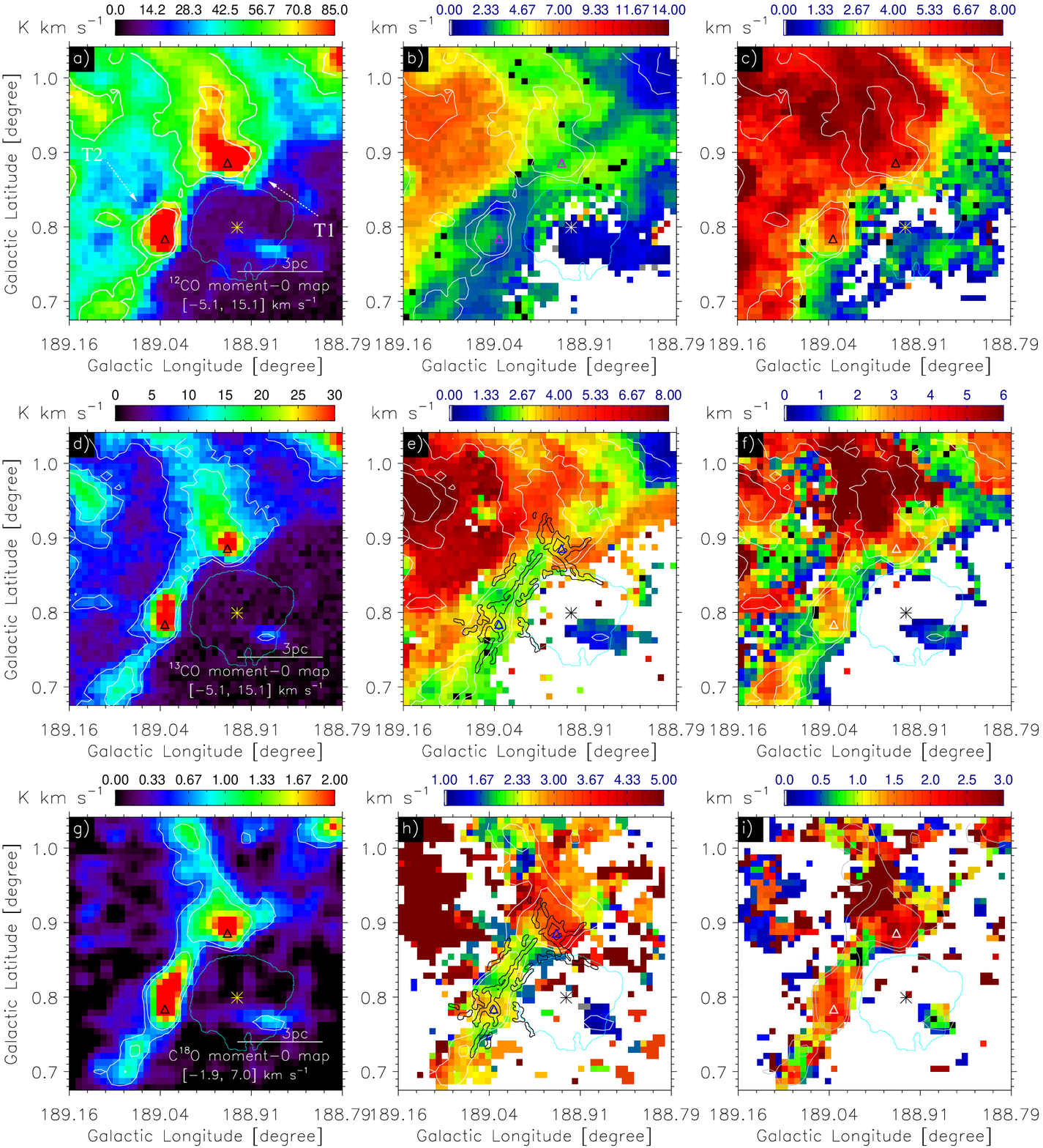}
\caption{Moment maps of CO molecular line data. Moment-0 maps of a) {\twco}; d) {\tco}; g) {\cetno}. Moment-1 maps of b) {\twco}; e) {\tco}; h) {\cetno}. Linewidth (FWHM) maps of c) {\twco}; f) {\tco}; i) {\cetno}. All the moment maps of {\twco} (i.e., the top row) are overplotted with {\twco} integrated emission contours at 30\% and 40\% of the peak value (i.e., {\simi}174~{\kkps}). Similarly, the moment maps of {\tco} (i.e., the middle row) and {\cetno} (i.e., the bottom row) are overplotted with their integrated emission contours at 20\% and 30\% of corresponding peak values (i.e., {\simi}43\,{\kkps} and {\simi}3.3\,{\kkps}, respectively). The moment-1 maps of {\tco} and {\cetno} emission are overlaid with the filament skeletons in black. The velocity ranges of integration are given in each moment-0 map. In all panels, the NVSS 1.4\,GHz radio continuum contour at 1.35\,mJy\,beam$^{-1}$ is also shown (in cyan), and other symbols are the same as shown in Figure~\ref{fig1}a.}
\label{fig5}
\end{figure*}
\begin{figure*}
\includegraphics[width=0.8\textwidth]{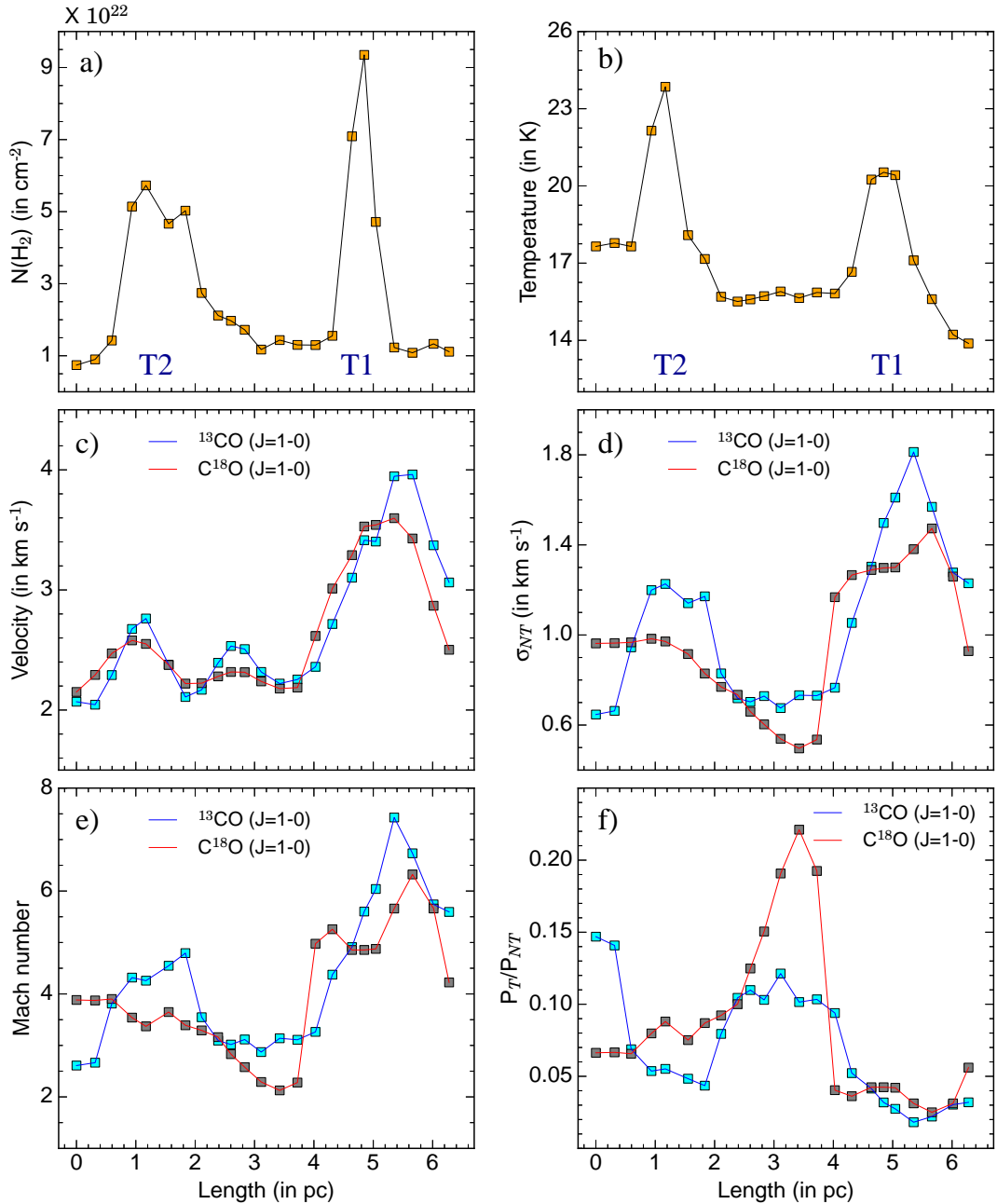}
\caption{Distribution of a) average column density, b) average dust temperature, c) radial velocity, d) non-thermal velocity dispersion, e) Mach number, and f) ratio of thermal to non-thermal gas pressure for the circular regions shown in Figure~\ref{fig4}e. In panels ``c--f'', physical parameters are shown for {\tco} and {\cetno} line data.}
\label{fig6}
\end{figure*}

\begin{figure*}
\includegraphics[width= \textwidth]{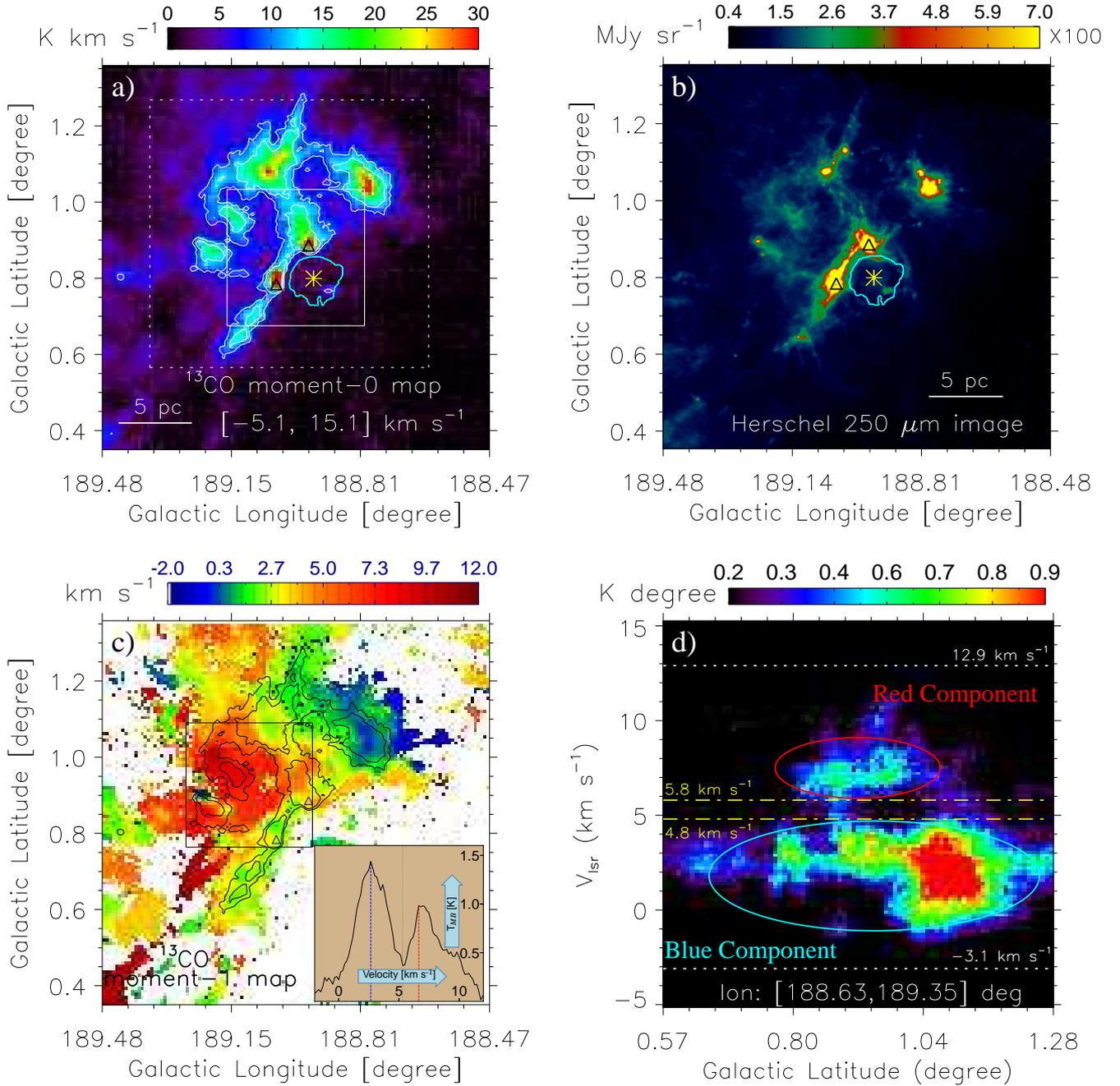}
\caption{ a) Moment-0 map of {\tco} emission toward our target sites for an extended area of about 60\rlap{\am} $\times$ 60\rlap{\am} and a velocity integration range [-5.1, 15.1]\,{\kps}. The contours are at the 20\% and 30\% of the peak value (i.e., {\simi}43\,{\kkps}). A scale bar of 5\,pc is shown along with other symbols identical to Figure~\ref{fig5}. The white rectangle (solid) presents the area of the molecular data shown in Figure~\ref{fig5}. The white rectangle (dotted) indicates the area of the molecular data utilized for the Galactic latitude--velocity diagram shown in Figure~\ref{fig7}d. b) {\it Herschel} 250\,{\micro} dust continuum image for the same region as Figure~\ref{fig7}a. c) Moment-1 map of {\tco} data with identical contours as in Figure~\ref{fig7}a. The black rectangle presents the area utilized to extract the average intensity profile, which is shown in the inset. d) Galactic latitude--velocity diagram for Galactic longitude integration range [188.63, 189.35]\,degree. The dashed and dotted horizontal lines present different V$_{\mathrm{lsr}}$ values, as the figure mentions. The cyan and red ellipses show the blue-shifted and red-shifted cloud components, respectively.}
\label{fig7}
\end{figure*}
\begin{figure*}
\includegraphics[width= \textwidth]{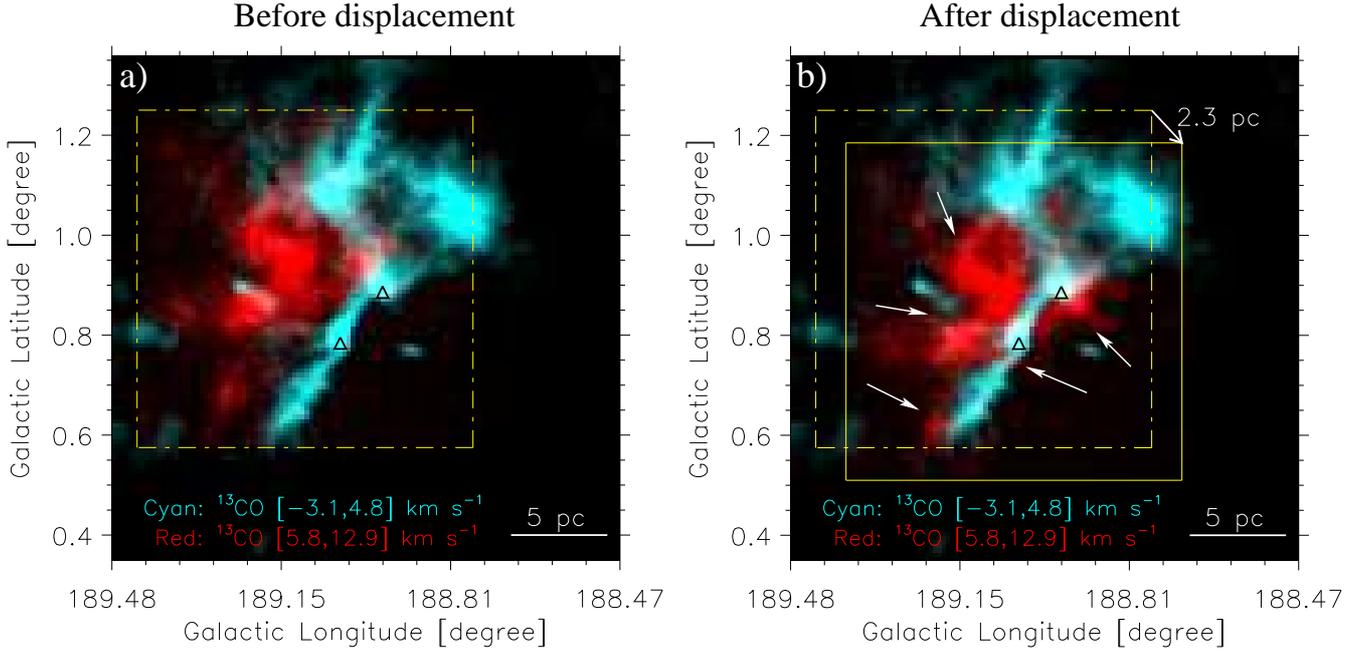}
\caption{ a) A two-color composite image shows the spatial distribution of the blue and red cloud components. The moment-0 map for the blue component is displayed in cyan on the linear scale from 2 to 14\,{\kkps}. Similarly, the moment-0 map for the red component is shown in red, ranging from 0 to 8\,{\kkps} on the linear scale within the yellow dashed rectangle. b) Spatial distribution of the blue and red cloud components with a spatial shift of about 2.3\,pc. The dashed and solid yellow rectangle presents the initial and ﬁnal positions of the red cloud component, respectively. The white arrows overlaid indicate toward the complementary distribution between the cloud components. The velocity integration ranges for the blue and red cloud components are specified in both the panels. The scale bar and other symbols are identical to Figure~\ref{fig7}a.}
\label{fig8}
\end{figure*}
%
%
%
\bibliographystyle{mnras}
\bibliography{reference} 
\appendix
\section{Plane of sky magnetic field in our target area}
\label{sec:posbfield} 
\begin{figure}
\includegraphics[width= 8 cm]{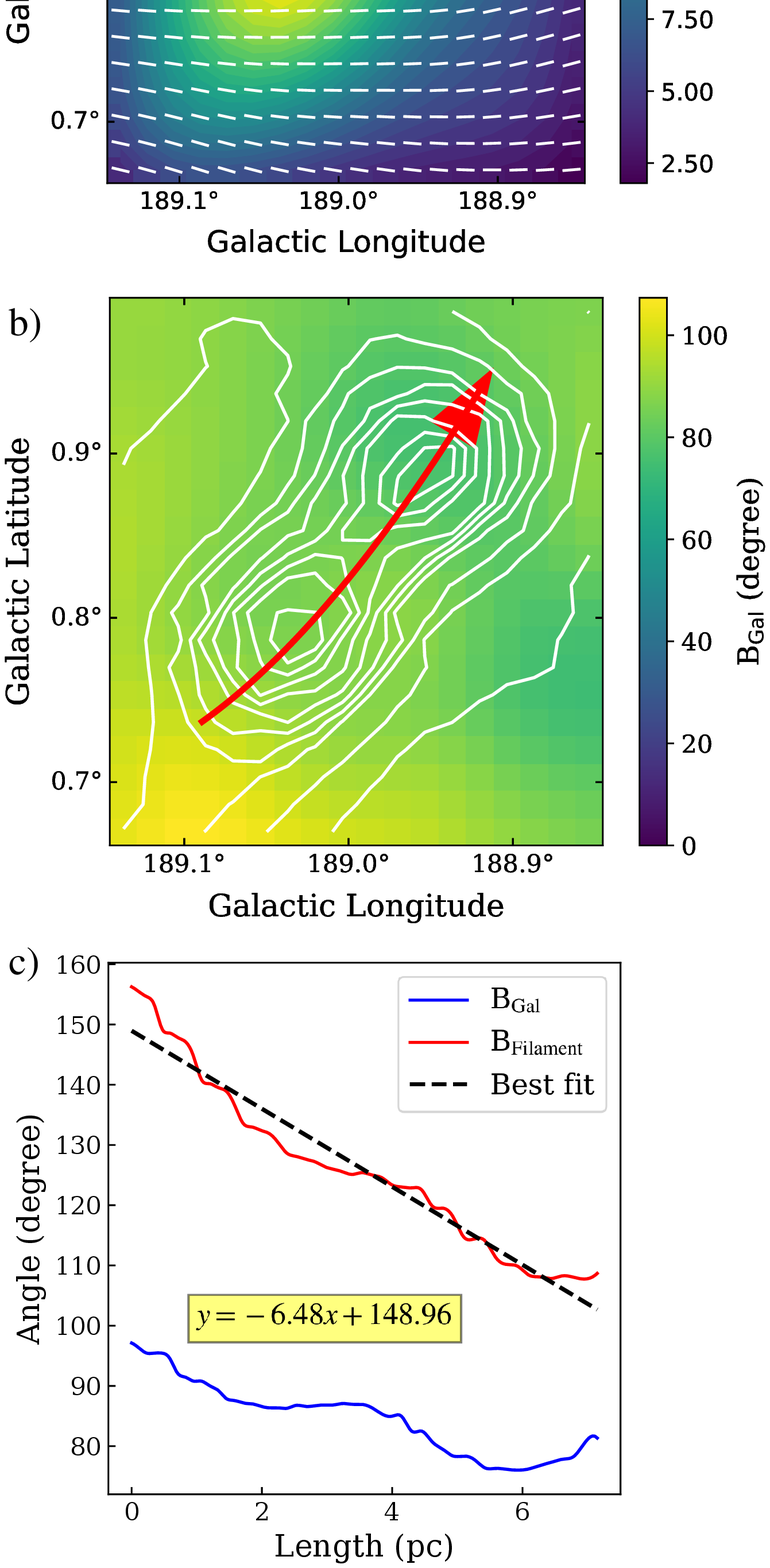}
\caption{a) The {\it Planck} 353\,GHz map of our target area (see the white dotted rectangle in Figure~\ref{fig8}a) overlaid by the POS magnetic field pseudo-vectors. b) Spatial distribution of the POS magnetic field position angle measured from Galactic north to the anticlockwise direction (B$_{\rm Gal}$). The overlaid contours indicate the {\it Planck} 353\,GHz intensity ranging from 4.6 to 27.6\,MJy\,sr$^{-1}$ in linear order. c) Distribution of B$_{\rm Gal}$ and B$_{\rm Filament}$ along the red curved arrow shown in panel ``b''. B$_{\rm Filament}$ is the magnetic ﬁeld position angle measured from ﬁlament's major axis (see text in Appendix~\ref{sec:posbfield} for more details). A best-fitted line is displayed and labeled for the B$_{\rm Filament}$ distribution.}
\label{fig9}
\end{figure}

We derived the plane-of-sky (POS) magnetic field position angles using the {\it Planck} 353\,GHz Stokes I, Q, and U images. We first estimated the linear polarization angles (PAs) of dust emission in Galactic coordinates using the conventional relation of $\theta_{\rm GAL}=0.5\times{\rm arctan2(-U, Q)}$. The negative sign in U is used to follow the IAU convention \citep[see more details in][]{Planck_2015_apr} and a two-argument function arctan2 is used to avoid the $\pi$-ambiguity in the estimation of PAs. The magnetic field orientations (B$_{\rm Gal}$) were then computed by adding 90$\degr$ in the electric field PAs \citep[e.g.,][]{Planck_2016_feb,Planck_2016_sep}. B$_{\rm Gal}$ is measured from the Galactic north to the anticlockwise direction (i.e., toward east).

Figure~\ref{fig9}a displays the {\it Planck} 353\,GHz image of our target site. The distribution of the POS magnetic field is shown by the overlaid pseudo-vectors in Figure~\ref{fig9}a. The magnetic field direction is nearly perpendicular to the filament, consistent with the observations of other targets \citep[e.g.,][]{Palmeirim_2013,Planck_2016_feb_A138,Cox_2016}. The spatial distribution of B$_{\rm Gal}$ is shown in Figure~\ref{fig9}b. We have also studied the variation of magnetic field position angle with respect to the filament's major axis (i.e., B$_{\rm Filament}$). The variation of B$_{\rm Gal}$ and B$_{\rm Filament}$ toward our target area is shown in Figure~\ref{fig9}c. A linear trend in B$_{\rm Filament}$ along filament's major axis hint at longitudinal mass flow from the edges toward the central part \citep{Dewangan_2023}. However, new high-resolution dust polarization data toward our target area can shed more light on the role of the magnetic field on the observed morphology. 

\section{A zoomed-in view of massive star-forming region T1}
\label{sec:ALMA_res} 
\begin{figure*}
\includegraphics[width= \textwidth]{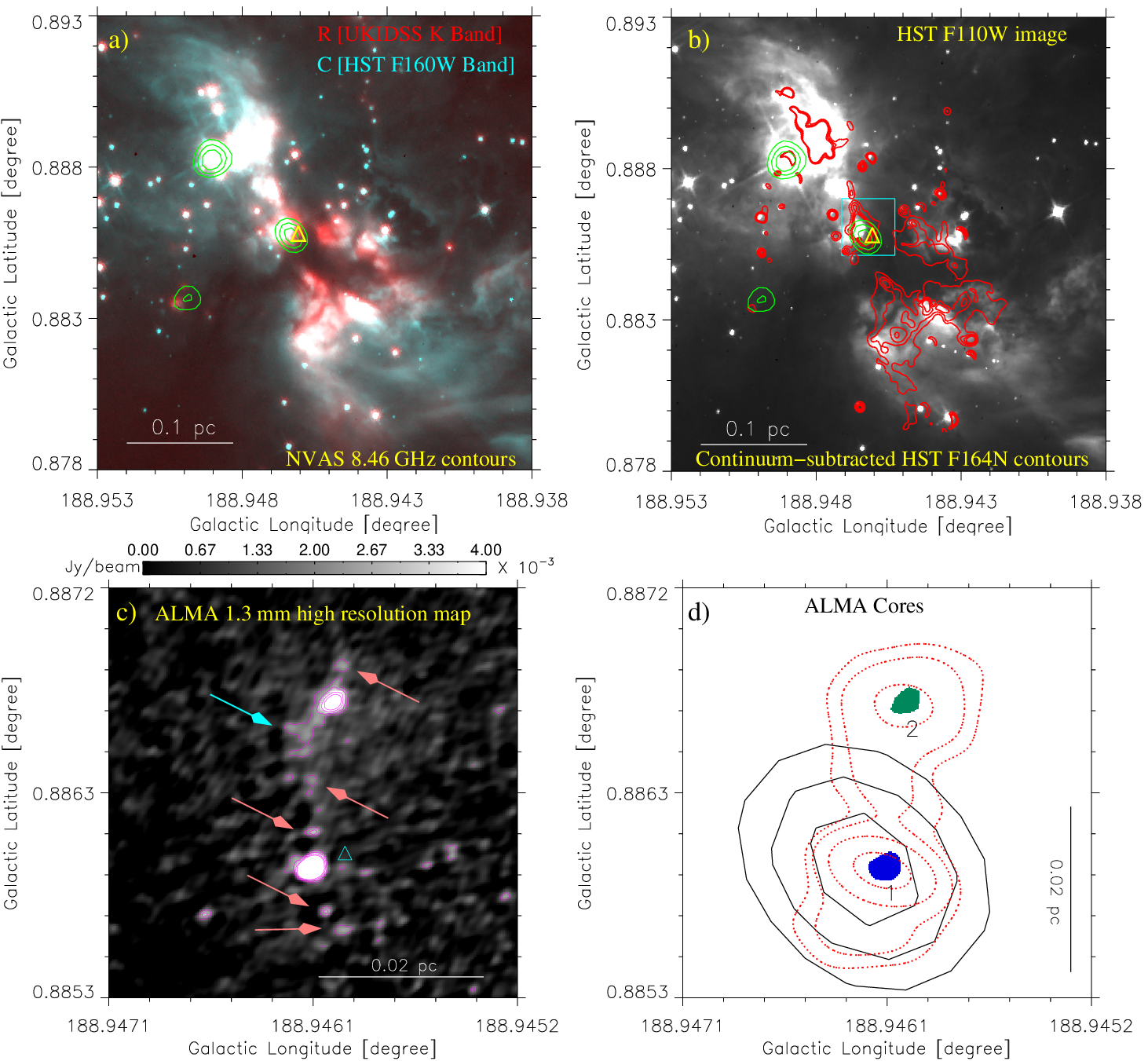}
\caption{a) A zoomed-in view of the site AFGL\,5180 (for a rectangular area shown in Figure~\ref{fig4}b) using a two-color composite image (UKIDSS K-band image in red + HST F160W band image in cyan). The NVAS 8.46\,GHz radio continuum emission contours (in green) are also overlaid on the color composite image. b) The panel presents the HST F110W wide band image overplotted with the continuum-subtracted [Fe\,{\sc ii}] emission contours (at 1.64\,{\micro}; in red) and the NVAS 8.46\,GHz radio continuum emission contours (in green). c) The panel presents the ALMA 1.3\,mm dust continuum map (resolution $\sim$0\rlap{\as}. 18 $\times$ 0\rlap{\as}. 28) of an area highlighted by a rectangle in Figure~\ref{fig10}b. The 1.3\,mm dust continuum emission contours are shown  at values 4$\sigma$, 6$\sigma$, 9$\sigma$ and 12$\sigma$, where 1$\sigma$ = 0.45\,mJy\,beam$^{-1}$. The arrows indicate possible low-mass cores and an envelope-like feature (see Appendix~\ref{sec:ALMA_res} for more details). 
d) The panel shows the positions of two cores (i.e., 1 and 2) detected in the ALMA 1.3\,mm dust continuum map (resolution $\sim$0\rlap{\as}. 18 $\times$ 0\rlap{\as}. 28). The NVAS 8.46\,GHz radio continuum emission contours are also plotted in black (see Figure~\ref{fig10}a). The dotted contours in red show the 1.3\,mm dust continuum emission (resolution $\sim$0\rlap{\as}. 63 $\times$ 1\rlap{\as}. 23) with the levels of 8, 14, 20, 26, and 32\,mJy\,beam$^{-1}$.
In panels ``a'', ``b'' and ``c'', the triangle indicates the position of the 6.7\,GHz MME, and the radio continuum contours are plotted with the levels of 120, 160, and 200\,$\mu$Jy\,beam{$^{-1}$}, where 1$\sigma$ = 36.8\,$\mu$Jy\,beam{$^{-1}$}. A scale bar of 0.1\,pc and 0.02\,pc is shown in panels (``a'', ``b'') and (``c'', ``d''), respectively.}
\label{fig10}
\end{figure*}

UKIDSS and HST high-resolution NIR images allow us to examine a zoom-in view of T1 (see the magenta rectangle in Figure~\ref{fig4}b). In Figure~\ref{fig10}a, we present a two-color composite image (based on UKIDSS K band image in red and HST F160W band image in cyan), overlaid with the NVAS 8.46\,GHz radio continuum emission contours in green, showing the presence of three compact radio sources toward the site AFGL\,5180. Ionized areas smaller than 0.05\,pc in size are previously referred to as the HC {\htwo} regions \citep{Yang_2021}, which are younger than the UC {\htwo} regions. Toward AFGL\,5180 one of the radio continuum sources is exclusively associated with the Class II 6.7\,GHz MME. Other radio sources lack detection of the 6.7\,GHz MME, and one of them (at the top) spatially coexists with a B4V-B8V spectral type star \citep{Vasyunina_2010}.

Following the method described in \citet{Long_2020}, the continuum-subtracted [Fe\,{\sc ii}] image at 1.64\,{\micro} is produced using the HST F164N and F160W band images. The shock-excited materials due to outflows and jets are the sources of [Fe\,{\sc ii}] emissions in star-forming regions \citep{Fedriani_2019}. In the direction of T1, we have displayed the continuum-subtracted [Fe\,{\sc ii}] emission at 1.64\,{\micro} with the help of red contours on the HST F110W band image (see Figure~\ref{fig10}b). Interestingly, the 6.7\,GHz MME and the NVAS 8.46\,GHz radio continuum emission are detected at the center of the bipolar morphology (or the bipolar outflow).

The area highlighted by a cyan box in Figure~\ref{fig10}b is further zoomed-in using the ALMA 1.3\,mm dust continuum emission (resolution $\sim$0\rlap{\as}. 18 $\times$ 0\rlap{\as}. 28), and is presented in Figure~\ref{fig10}c. The presence of at least two cores (i.e., core-1 and core-2) is evident in Figure~\ref{fig10}c. The cyan arrow indicates an extended dust emission surrounding one of the cores. The pink arrows point out the candidate cores residing on the line joining the primary cores. Using the formula of \citet{Hildebrand_1983} \citep[see Equation~3 in][]{Mutie_2021}, we obtained the masses of the core-1 and core-2 to be about 5.4\,{\Msolar} and 1.3\,{\Msolar}, respectively. We have chosen 6$\sigma$ as the contour value to evaluate the total dust emissions using the {\it clumpfind}. The distance and the temperature of the source are taken to be 1.5\,kpc and 25\,K, respectively. The gas-to-dust ratio and the opacity value are adopted to be 100 and 0.33\,cm$^2$\,g$^{-1}$, respectively \citep{Weingartner_2001,Mutie_2021}. Additionally, the mass of the elongated envelope around core~2 is determined to be about 1.1\,{\Msolar}. Earlier, \cite{Mutie_2021} reported the mass of the core-1 and core-2 to be 8.2\,{\Msolar} and 4.8\,{\Msolar}, which are different from our estimations. The difference with our estimated mass values mainly arises due to the use of the different values of distance and temperature. 
The core-1 and core-2 traced in the ALMA 1.3\,mm continuum emission map (resolution $\sim$0\rlap{\as}. 18 $\times$ 0\rlap{\as}. 28) are presented in Figure~\ref{fig10}d. We have also shown the 1.3\,mm continuum emission (resolution $\sim$0\rlap{\as}. 63 $\times$ 1\rlap{\as}. 23) contours (in red), the NVAS 8.46\,GHz radio continuum emission contours (in black), and the position of the 6.7\,GHz MME in Figure~\ref{fig10}d. In Figure~\ref{fig10}d, we find that the core-1 is associated with the 6.7\,GHz MME and the radio continuum emission. The core-1 (mass $\sim$5.4\,{\Msolar}) is connected with the core-2 (mass $\sim$1.3\,{\Msolar}) by the dust emission (extent $<$ 5000\,AU), showing a dumbbell-like configuration at small scale. Such configuration is traced within the central hub of the HFS in AFGL\,5180. One can also notice that no radio continuum emission is seen toward the core-2. 
%
\end{document}